\documentclass[12pt]{article}

\usepackage{verbatim}
\usepackage{amsmath}
\usepackage{amsfonts}
\usepackage{eucal}
\usepackage{latexsym}
\usepackage{epsfig}
\usepackage{color}
\if@twoside\oddsidemargin25mm
\else
\renewcommand{\d}{\mathrm{d}}

\newcommand{\Ra}{\Rightarrow}

\DeclareMathSymbol{\mg}{\mathrel}{symbols}{"1D}

\newcommand{\dga}{{\dot{\alpha}}}
\newcommand{\dgb}{{\dot{\beta}}}

\newcommand{\ga}{\alpha}
\newcommand{\gb}{\beta}

\newcommand{\gd}{\delta}
\renewcommand{\ge}{\epsilon}

\newcommand{\gf}{\phi}
\newcommand{\gvf}{\varphi}

\newcommand{\gx}{\xi}
\newcommand{\gm}{\mu}

\newcommand{\gk}{\kappa}
\newcommand{\gl}{\lambda}

\newcommand{\gth}{\theta}

\newcommand{\gs}{\sigma}

\newcommand{\go}{\omega}

\newcommand{\gp}{\pi}

\newcommand{\get}{\eta}

\newcommand{\gG}{\Gamma}

\newcommand{\gF}{\Phi}

\newcommand{\gL}{\Lambda}

\newcommand{\gO}{\Omega}

\newcommand{\gPs}{\Psi}

\newcommand{\cA}{{\cal A}}
\newcommand{\cB}{{\cal B}}
\newcommand{\cC}{{\cal C}}
\newcommand{\cD}{{\cal D}}

\newcommand{\cF}{{\cal F}}

\newcommand{\cH}{{\cal H}}

\newcommand{\cK}{{\cal K}}
\newcommand{\cL}{{\cal L}}

\newcommand{\cP}{{\cal P}}

\newcommand{\cS}{{\cal S}}
\newcommand{\cT}{{\cal T}}
\newcommand{\cU}{{\cal U}}
\newcommand{\cV}{{\cal V}}
\newcommand{\cW}{{\cal W}}

\newcommand{\tr}{\text{tr}}

\newcommand{\der}{\partial}

\newcommand{\dsp}{\displaystyle}

\newcommand{\labl}[1]{\label{#1}}
\newcommand{\beq}{\begin{equation}}
\newcommand{\eeq}{\end{equation}}
\newcommand{\barr}{\begin{array}}
\newcommand{\earr}{\end{array}}
\newcommand{\equ}[1]{\begin{gather} #1 \end{gather}}
\newcommand{\equa}[1]{\begin{align} #1 \end{align}}

\newcommand{\arry}[2]{\begin{array}{#1} #2 \end{array}}

\newcommand{\pmtrx}[1]{\begin{pmatrix} #1 \end{pmatrix}}

\newcommand{\non}{\nonumber}
\newcounter{oldcounter}

\newcommand{\bD}{{\bar D}}

\newcommand{\bF}{{\bar F}}

\newcommand{\bS}{{\bar S}}

\newcommand{\bW}{{\bar W}}

\newcommand{\bY}{{\bar Y}}

\newcommand{\bgvf}{{\bar\varphi}}

\newcommand{\bgth}{{\bar\theta}}

\newcommand{\bgF}{{\bar\Phi}}

\newcommand{\bgL}{{\bar\Lambda}}

\newcommand{\bgPs}{{\bar\Psi}}

\newcommand{\bcF}{{\bar{\cal F}}}

\newcommand{\bcS}{{\bar{\cal S}}}
\newcommand{\bcT}{{\bar{\cal T}}}

\newcommand{\Intr}{\mathbb{Z}}

\newcommand{\ba}[2]{\[\begin{array}{#2}\label{#1}}
\newcommand{\ea}{\end{array}\]}
\newcommand{\be}{\begin{equation}}
\newcommand{\ee}{\end{equation}}
\newcommand{\bea}{\begin{eqnarray}}
\newcommand{\eea}{\end{eqnarray}}

\newcommand{\U}[1]{\mathrm{U(#1)}}

\begin{document}

\thispagestyle{empty}

\begin{flushright}
CPHT RR 010.0304 \\
LPT-ORSAY 04-28\\
FTPI-MINN-04/12 \\
UMN-TH-2301/04  \\      
hep-th/0404094
\end{flushright}
\vskip 2 cm
\begin{center}
{\Large {\bf 
Vector/tensor duality in the five dimensional 
\\[1ex]
supersymmetric Green--Schwarz mechanism 
}
}
\\[0pt]
\vspace{1.23cm}
{
{\sc Emilian Dudas$^{a,b,}$\footnote{
\ E-mail: emilian.dudas@th.u-psud.fr}}, 
{\sc Tony Gherghetta$^{c,}$\footnote{
\ E-mail: tgher@physics.umn.edu}},
{\sc Stefan Groot Nibbelink$^{d,}$\footnote{
\ E-mail: nibbelin@physics.umn.edu\\[-2ex]}}
\bigskip }\\[0pt]
\setcounter{footnote}{0}
\renewcommand{\thefootnote}{\fnsymbol{footnote}}
\vspace{0.23cm}
${}^a$ {\it\small
Centre de Physique Th{\'e}orique\footnote{\ Unit{\'e} mixte du CNRS et de l'EP,
UMR 7644.} , Ecole Polytechnique, 
F--91128 Palaiseau, France 
\\
${}^b$ Laboratoire de Physique Th{\'e}orique\footnote{\ Unit{\'e} Mixte de
Recherche du CNRS (UMR 8627).}, Universit{\'e} de Paris-Sud, F--91405 Orsay Cedex, 
France}
\\
${}^c$ {\it\small
School of Physics \& Astronomy,
University of Minnesota,
Minneapolis, MN 55455, USA}
\\
${}^d$ {\it\small
William\,I.\,Fine\,Theoretical\,Physics\,Institute,
University of Minnesota,\,
Minneapolis,\,MN 55455,\,USA}

\bigskip
\end{center}
{\small ABSTRACT:} 
The five dimensional version of the Green--Schwarz mechanism can be
invoked to cancel $\U{1}$ anomalies on the boundaries of brane world
models. In five dimensions there are two dual descriptions that employ
either a two--form tensor field or a vector field. We present the
supersymmetric extensions of these dual theories using four dimensional
$N\!=\!1$ superspace. For the supersymmetrization of the five dimensional
Chern--Simons three form this requires the introduction of a new chiral
Chern--Simons multiplet.  We derive the supersymmetric vector/tensor
duality relations and show that not only is the usual one/two--form
duality modified, but that there is also an interesting duality relation
between the scalar components. Furthermore, the vector formulation always
contains singular boundary mass terms which are absent in the tensor
formulation. This apparent inconsistency is resolved by showing that in
either formulation the four dimensional anomalous U(1) mass spectrum is
identical, with the lowest lying Kaluza--Klein mode generically obtaining 
a finite nonzero mass.

\newpage 
\setcounter{footnote}{0}
\renewcommand{\thefootnote}{\arabic{footnote}}
\setcounter{page}{1}

\section{Introduction}
\labl{sc:intro}

The quantum consistency of gauge theories coupled to matter, specifically
the absence of anomalies, has proven to be one of the most important guiding
principles for model building beyond the Standard Model. Therefore not
surprisingly there have been many investigations \cite{ch,hw1,hw2,bddr,Arkani-Hamed:2001is,Scrucca:2001eb,Pilo:2002hu,Barbieri:2002ic,GrootNibbelink:2002qp}
of anomalies in brane world models, or in particular orbifold
theories. The general outcome of these analyses is that the anomalies
localize on even dimensional hypersurfaces and are
determined by the local fermionic spectrum of the theory. (This
conclusion also remains true in the context of warped compactifications 
\cite{Gherghetta:2002nq,Hirayama:2003kk}). In odd dimensions
anomalies do not need to cancel locally, since as long as they cancel
globally, a bulk Chern--Simons term can be used to ensure that the
theory is gauge invariant at the quantum level \cite{Arkani-Hamed:2001is}. 
Therefore, an interesting question is whether it is possible to have even
more general localized anomalies that can still lead to consistent 
theories.

One of the major breakthroughs in the development of string theory was
the realization of Green and Schwarz \cite{Green:1984sg} that so--called
factorisable anomalies can be compensated by anomalous variations of
anti--symmetric tensor fields. Studies of heterotic string
compactifications to four dimensions has revealed that a four
dimensional version of the Green--Schwarz mechanism is relevant for
phenomenological string model building \cite{dsw}.  This mechanism 
simultaneously cancels both pure and mixed $\U{1}$ anomalies provided 
that the spectrum satisfies a specific universality condition
\cite{Schellekens:1986xh,Kobayashi:1996pb}. In particular, there will
be a mixed gravitational--gauge anomaly which in the $N\!=\!1$
supersymmetric context is directly related to one--loop induced
Fayet--Iliopoulos terms \cite{Fayet:jb,Fischler:1981zk} that has 
been confirmed in the string theory context
\cite{dsw,Atick:1987gy,Dine:1987gj}\footnote{
In the generic context of Type II and orientifold
models the situation is quite different. The Fayet--Iliopoulos terms 
generically have a tree--level origin, whereas the Green-Schwarz field is a
localized brane field arising from the twisted sector of an orbifold
compactification. Since the field theory description of this
phenomenon presents no subtlety, we do not consider it here in detail.  
There are however some instances in orientifold models (e.g., in
models with branes at angles), in which the Green--Schwarz field is a bulk 
field and the Fayet--Iliopoulos term is one--loop induced.}.  
The local version of the Green--Schwarz mechanism on (strongly
coupled) heterotic orbifolds has been investigated in
\cite{Gmeiner:2002es,GrootNibbelink:2003gb, Nibbelink:2003rc}, and 
the corresponding localized Fayet--Iliopoulos tadpoles were computed in 
\cite{GrootNibbelink:2003zm}.

This provides motivation to investigate how the local Green--Schwarz 
mechanism can be implemented in a five dimensional setting. In odd 
dimensions the prime example of the Green--Schwarz--like mechanism 
is given by eleven dimensional Horava--Witten theory \cite{hw1,hw2,bddr}. An
interesting aspect of the Green--Schwarz theory is that it allows for
dual descriptions \cite{Gates:dm}. In five dimensions the Green--Schwarz interaction 
can be described by two equivalent formulations using either a
rank two tensor or a vector.  (We do not consider the trivial 
possibility that boundary axion states could also cancel the anomalies.) 
As we will show this duality indeed occurs
in our five dimensional theory even though the two descriptions
are not always manifestly equivalent. For example, the vector 
formulation contains singular boundary mass terms for the anomalous 
$\U{1}$ gauge field which are absent when the two--form formulation 
is used. However, the proper inclusion of mixing terms in both the
vector and the tensor formulations resolves this apparent
inconsistency. In fact, the consistency of the duality provides confirmation 
of the $\gd(0)$ regularization proposed in Ref.~\cite{mp}.

Supersymmetry plays a pivotal role in string theory, and therefore
it is important to construct a manifestly supersymmetric extension of
this mechanism in the context of (global) five dimensional
supersymmetry. An elegant and simple way to do this is to 
employ four dimensional $N\!=\!1$ superspace techniques in five dimensions
\cite{Marcus:1983wb,Arkani-Hamed:2001tb,Hebecker:2001ke,Linch:2002wg}. In
particular, for the tensor formulation we use the five dimensional 
tensor multiplet and find that contrary to the conventional four 
dimensional case, the $N\!=\!1$ linear supermultiplet is insufficient 
to contain all the bosonic components resulting from the dimensional 
reduction, and it cannot reproduce $\der_5 \cB_{mn}$ terms that are present 
in the five dimensional action. In addition the Green--Schwarz theory contains
Chern--Simons interactions. The five dimensional completion 
of the four dimensional Chern--Simons (three--form) superfield
will require an additional chiral Chern--Simons superfield. 

The supersymmetric version of the vector/tensor duality is investigated
in detail and leads to new duality relations. In particular the usual
non--supersymmetric duality relation between one-- and two--forms is modified,
while a new relation appears for the scalar field components of the 
supermultiplets. This relation is closely related to a duality for
scalars in five dimensional supersymmetric gauge 
theories~\cite{Seiberg:1996bd}. Although we are mainly concerned 
about a single $\U{1}_A$ anomalous gauge group, it is also possible to 
consider the generalization to multiple $\U{1}_A$ gauge groups. In fact
these results can be conveniently written in terms of a prepotential, 
and an interesting feature of this generalization is that so--called 
generalized Chern--Simons interactions \cite{wlp,akt,aafl,wst,dfp,afl} will 
naturally appear.

The presence of anomalous symmetries in the four dimensional version of
the Green--Schwarz mechanism plays an important role in
phenomenological applications of various compactifications of string 
theories. Similarly, we expect that the five dimensional formalism 
developed here could be useful for the fermion mass hierarchy 
and supersymmetry breaking in the context of models with large extra
dimensions and a low fundamental scale.

The plan of this paper is as follows: The paper is divided
into two main parts. First, in section
\ref{sc:GSbosonic} we give a detailed account of the vector/tensor 
duality in five dimensions and explain its role in the five dimensional 
Green--Schwarz mechanism. This duality will then be used
to resolve the seemingly ill--defined boundary $\gd(0)$ mass terms. 
In the second part of the paper we present the supersymmetric dual
formulations of the Green--Schwarz theory using four dimensional
superspace techniques. In section \ref{sc:SusyVT} we define
the vector and tensor multiplet in this formalism and 
then proceed to the supersymmetric version of the
vector/tensor multiplet duality in five dimensions. 
Next, in section \ref{sc:SusyGS} we systematically supersymmetrize all
elements on the Green--Schwarz theory, including the
Chern--Simons interactions. We derive the superfield duality relations 
and use them to consistently check that the dual supersymmetric descriptions
are equivalent. For example, even in the supersymmetric context singular 
boundary terms are absent in the tensor multiplet formulation. 
Finally, we briefly discuss various phenomenological applications of the
supersymmetric local Green--Schwarz theory in section \ref{sc:Pheno}
and summarize our main results in the Conclusions. Various details 
required for the systematic construction of the Chern--Simons superfields
are given in Appendix \ref{sc:SCSdefs}, while in Appendix \ref{sc:MultiU1} 
we present the extension  of the supersymmetric Green--Schwarz theory 
to the case of multiple  anomalous $\U{1}_A$ superfields.

\section{Bosonic Green--Schwarz mechanism in five dimensions}
\labl{sc:GSbosonic}

We begin with a review of the Green--Schwarz anomaly
cancellation mechanism in the context of a five dimensional gauge
theory on the interval $S^1/\Intr_2$. The spacetime coordinates are
denoted by $x^M =(x^0, \ldots x^3, x^5 = y)$, with 
$0 \leq y \leq \pi$ (where we have set the radius $R$ of the extra
dimension and the five dimensional fundamental scale to unity). For 
simplicity we will consider
the gauge potential $A_M$, $M = 0, \ldots 3, 5$, that corresponds to a
single $\U{1}_A$ gauge group with field strength $F_{MN}$.  
As is well--known bulk fermions and brane chiral
fermions give rise to gauge anomalies localized on the boundaries, as
was shown by direct calculations 
\cite{bddr,Arkani-Hamed:2001is,Scrucca:2001eb,Barbieri:2002ic}. 
When the anomalies on the two boundaries are equal and
opposite, they can be cancelled by a five dimensional Chern--Simons 
interaction. However this is a very special form for the anomaly and
more general boundary anomalies can instead be
cancelled by invoking the Green--Schwarz mechanism~\cite{Green:1984sg}.
In five dimensions the Green--Schwarz mechanism can be
described by either a rank two tensor $\cB_{MN}$ or a vector
$\cA_M$ that have field strengths $\cH_{MNP}$ and $\cF_{MN}$, respectively.

Let us emphasize our notational conventions: We use calligraphic
letters $\cA_1$, $\cB_2$ and $\cF_2$, $\cH_3$ to denote the
non-anomalous Green--Schwarz one and two--forms, and their
corresponding field strengths. The anomalous $\U{1}_A$ fields will
instead be denoted by the standard italic letters $A,F,\ldots$.   
In the following sections this notation will be extended to the
supersymmetric case as well. The superfields corresponding to the
Green--Schwarz multiplets are denoted by calligraphic letters
($\cT_\ga, \cU, \cV, \cS$), while standard italic letters ($V, S$) are
reserved for the anomalous $\U{1}_A$ vector multiplet.


\subsection{Anomalous $\boldsymbol{\U{1}_A}$ gauge field in the bulk}
\labl{sc:anomGauge}

The gauge field $A_M$ in the bulk has the standard Yang--Mills action 
\equ{
S_{YM} = \int \frac 12 *\! F_2 F_2 = \int  \frac 12 *\! \d A_1 \d A_1 
= \ - \ \int \d^5 x\,  \frac 14 F_{MN} F^{MN}~,
\labl{YM}
}
where we have introduced form notation for brevity and compliance with
the standard literature on anomalies \cite{Zumino:1984rz,gsw2}.
The field strength two--form 
$F_2 = \frac 12 F_{MN} \d x^M \d x^N$ is defined as the exterior
derivative $\d = \d x^M \der_M$ of the one--form $A_1 = A_M \d x^M$. 
In components  $F_2 = \d A_1$ is written as $F_{MN} = \der_M A_N - 
\der_N A_M$. The differentials $\d x^M$ are understood to be multiplied 
by the anti--commuting wedge product: $\d x^M \d x^N = - \d x^N \d x^M$. 
On a two--form the Hodge star $*$ is defined as 
\equ{
*(\d x^M \d x^N) = \frac 1{3!} 
\ge^{MNPQR} \get^{\;}_{PS} \get^{\;}_{QT} \get^{\;}_{RU} \d x^S \d x^T \d x^U~, 
\qquad 
*(*(\d x^M \d x^M)) = - \d x^M \d x^N~,
}
where $\ge^{MNPQR}$ is the totally anti--symmetric
epsilon tensor in five dimensions with $\ge^{01235} = 1$. 
Note that the sign of a two--form changes when the Hodge star is applied 
twice, due to the signature of the Minkowski metric $\get^{\;}_{MN} =
\text{diag}(-1, 1,1,1,1)$. For more details see the textbook 
\cite{Nakahara:1990th}.

By assumption the effective action $\gG_{eff}(A)$, obtained by
integrating out all chiral brane and bulk fermions, is not invariant 
under the gauge transformation $\gd A_1 = \d \ga_0$. The gauge
parameter $\ga_0$ is a real scalar function (i.e.\ a zero--form). 
However, because of the Wess--Zumino consistency 
conditions \cite{Wess:1971yu}, and the fact that gauge anomalies do not
exist in five dimensions, we infer that the variation of the effective
action takes the form 
\equ{
\gd_{\ga_0} \gG_{eff}(A) = \sum_I \gx_I \int \ga_0 F_2^2 \gd(y - I) \d y 
= \sum_I \frac 12 \gx_I \int \ga_0 F_2^2 \Big|_I~, 
\labl{anomgGeff}
}
with 
\equ{
\gx_I = \frac 1{24 \gp^2} \Big( \tr\, q_I^3  + \frac 12 \tr\,q_b^3 \Big) 
= \frac k{192 \gp^2}  \Big( \tr\, q_I + \frac 12 \tr\, q_b \Big)~, 
\labl{FIcoef}
}
where $\tr \, q_I^3$ denotes the cubic sum of charges of the chiral
fermions at the boundary $I = 0, \gp$, and $\tr\, q_b^3$ that of the bulk
fermions. Note that the bulk fermions give equal anomalies on both boundaries
\cite{bddr, Arkani-Hamed:2001is} (assuming that the boundary conditions of
bulk fields are not twisted). With the notation~$|_I$ we emphasize that
all fields are to be evaluated at the boundary $y=I$. The factor of
$1/2$ in the last equation arises because the delta functions $\gd(y-I)$
are defined on the end points of the interval. Specifically,
$\int_0^\pi dy~\delta(y) f(y) = f(0)/2$, for an arbitrary function $f(y)$,
and similarly at $y=\pi$.

In this work we will not treat the cancellation of mixed gravitational
gauge anomalies via the Green--Schwarz mechanism in detail. However,
if the charges satisfy the ``universality'' relation, expressed
by the second equality in \eqref{FIcoef}, then the Green--Schwarz mechanism
can cancel both pure gauge and mixed gravitational anomalies
simultaneously. Such ``universality'' relations are well--known in
heterotic string models \cite{Schellekens:1986xh,Kobayashi:1996pb}.
The proportionality factor $k$ in (\ref{FIcoef}) is called the level
and depends on the normalization of the anomalous $\U{1}_A$.

In the Green--Schwarz anomaly cancellation mechanism a crucial role is
played by the Chern--Simons three--form  
\equ{
\go_3(A) = A_1 F_2~, 
\qquad 
\d \go_3(A) = F_2^2~, 
\labl{CS3form}
}
for the anomalous $\U{1}_A$ gauge potential one--form $A_1$. Under a
gauge transformation, the gauge variation of the Chern--Simons three
form is given by 
\equ{ 
\gd_{\ga_0} \go_3(A)  = \d \ga_0 F_2~. 
\labl{varCS3form}
}


\subsection{Vector/tensor duality in five dimensions}
 \labl{sc:VecTenDual}

In five dimensions a vector $\cA_M$ is dual to a two--form tensor
$\cB_{MN}$. (Aspects of the vector/tensor duality in
five dimensions were discussed in the past, see for example \cite{aft}, 
in the context of the Type II -- heterotic duality.)
A convenient starting point to describe this duality is to use the action 
\equ{
S_{TV} = \int 
\Big( \frac 12 * \widehat \cF_2 \widehat \cF_2 - \d \cB_2 \widehat \cF_2 \Big)~,
\labl{FormDual5D}
}
which contains the non--dynamical two forms $\widehat \cF_2$ and $\cB_2$. 
This action is invariant under gauge transformations of the two--form
$\cB_2$ given by
\equ{
\gd_{\gb_1} \cB_2 = \d \gb_1~, 
\qquad 
\gd_{\gb_1} \widehat\cF_2 = 0~.  
\labl{TwoFormGauge}
}
There are two possible equations of motion arising from the 
action (\ref{FormDual5D}).
First, the equation of motion of the two--form tensor
$\cB_2$ leads to the constraint $\d \widehat \cF_2 = 0$. This
equation can be  (locally)
solved by Poincar{\'e}'s lemma for a one--form gauge potential $\cA_1$
\equ{
\widehat \cF_2 = \cF_2 = \d \cA_1~, 
\qquad 
\gd_{\gb_0}\cA_1  = \d \gb_0~. 
\labl{gaugecA}
}
The gauge transformation of $\cA_1$ leaves its associated field 
strength $\cF_2 = \d \cA_1$ invariant. The hatted notation 
$\smash{\widehat \cF_2}$ for the non--dynamical two--form field might appear
somewhat redundant here, but we will see in the next subsection, where
we discuss the dual formulations of the Green--Schwarz 
mechanism, that the simple relation $\smash{\widehat \cF_2 = \cF_2}$ 
will be modified. Using this solution in the action
(\ref{FormDual5D}), leads to the standard kinetic Abelian Yang--Mills action
\equ{
S_V = \int \frac 12 * (\d \cA_1) \d \cA_1 = -  \int \d^5 x\ 
\!  \frac 14 \cF_{MN} \cF^{MN}~. 
\labl{AcVector}
}
On the other hand, since the equation of motion for
$\smash{\widehat\cF_2}$ is purely algebraic, it can be easily solved
to give $\smash{-\widehat \cF_2 = *\d \cB_2}$.  
Thus, the duality between $\cA_1$ and $\cB_2$ is established by the two
equations for $\smash{\widehat \cF_2}$, namely
\equ{
\cF_2 = \d \cA_1 = \widehat \cF_2 = -* \d \cB_2 = - * \cH_3~.
\labl{VTduality}
}  
Notice that this duality equation encodes the equations of motion for
$\cA_1$ and $\cB_2$: By applying $\d$ to this equation leads to 
$\d * \d \cB_2 = 0$, while acting with $\d * $ gives $\d * \d \cA_1=0$. 
In (\ref{VTduality}) we have introduced the three--form field strength 
$\cH_3 = \d \cB_2$ of the two--form $\cB_2$, which has the components 
\(
\cH_{MNP} = \der_M \cB_{NP} + \der_N \cB_{PM} 
+ \der_P \cB_{MN}~. 
\)
Substituting the solution for $\widehat\cF_2$ back into the action 
(\ref{FormDual5D}), we find the kinetic action for a rank two tensor 
\equ{
S_T = \int \frac 12 *(\d \cB_2) \d \cB_2 =  -  \int \d^5 x\ 
\! \frac 1{12} \cH_{MNP} \cH^{MNP}~. 
}
Hence we see that in this case by eliminating $\widehat \cF_2$ 
the two--form $\cB_2$ has become dynamical. Obviously, both the
anti--symmetric tensor and the vector formulations describes three
physical on--shell degrees of freedom.


\subsection{Dual formulations of the Green--Schwarz mechanism} 
\labl{sc:DualGS}

The vector/tensor action \eqref{FormDual5D}, described in the
previous subsection, is extended to the Green--Schwarz
action 
\equ{
S_{GS} = \int \frac 12  *\! F_2 F_2 + 
 \frac 12  *\! \widehat\cF_2 \widehat \cF_2 - \d \cB_2 \Big( 
\widehat \cF_2 - * \go_3(A) + X_2(A) 
\Big) + \frac 12 *\go_3(A) \go_3(A)~, 
\labl{GSdual}
}
with the Chern--Simons three--form $\go_3(A)$ given in
\eqref{CS3form}. We have also included the standard kinetic term
\eqref{YM}  of the gauge field one--form $A_1$; it will be a
``spectator'' as far as the duality is concerned. The two--form
$X_2(A)$ only has support at the end points of the interval    
\equ{
X_2(A) = \sum_I \gx_I \, A_1 \, \gd(y - I) \d y~.
} 
This action has been chosen such that it cancels the anomalous
variation of the effective action $\gG_{eff}$ obtained by integrating
out the brane and bulk fermions:  
\equ{
\gd_{\ga_0}\big( \gG_{eff} + S_{GS} \big) = 0~. 
}
Since the anomalous variation \eqref{anomgGeff} of the effective action
$\gG_{eff}$  does not contain the Hodge $*$--dualization, we infer that the
Green--Schwarz tensors $\cB_{MN}$ and $\smash{\widehat\cF_{MN}}$ 
transform (up to exact terms) as  
\equ{
\gd_{\ga_0} \cB_2 =  - \ga_0 F_2~,
\qquad 
\gd_{\ga_0} \widehat \cF_2 = - * \! \d (\gd_{\ga_0} \cB_2) = *( \d \ga_0 F_2)~.
\labl{GaugeTensors}
}
By a four dimensional partial integration it follows that the term
involving the gauge variation of $X_2$ is gauge invariant by itself, 
and therefore the gauge variation of the Green--Schwarz action is given by 
\equ{
\gd_{\ga_0} S_{GS} = \int - \d ( \gd_{\ga_0} \cB_2) X_2(A) 
= \int - \ga_{0} F_2^2 \gx_I \gd(y - I) \d y~. 
\labl{varGS} 
}
This variation indeed cancels the anomalous contribution of
$\gG_{eff}$ given in \eqref{anomgGeff}. 
It should be stressed that this calculation shows that the anomalous
variation is independent of 
which dual formulation of the theory is being used, since we did not yet
choose to eliminate either $\smash{\widehat \cF_2}$ or $\cB_2$ from the theory. 

As in the previous subsection we can go to the one-- or two--form
formulation of the Green--Schwarz mechanism by integrating out $\cB_2$ or 
$\smash{\widehat \cF_2}$. The equation of motion for $\cB_2$ is 
$ \d \big( \smash{\widehat\cF_2} - * \go_3(A) + X_2(A) \big) = 0$ which is solved by 
\equ{
\widehat \cF_2 = \d \cA_1 +  * \go_3(A) - X_2(A)~,
\labl{B2eom}
}
using the one--form $\cA_1$.  Clearly, the gauge transformation
\eqref{gaugecA} associated with this one--form is still there. But in
order for $\smash{\widehat \cF_2}$ to have the required gauge transformation 
\eqref{GaugeTensors} it follows that $\cA_1$ transforms under the
anomalous $\U{1}_A$ as 
\equ{
\gd_{\ga_0} \cA_1 = \sum \gx_I \ga_0 \gd(y-I) \d y~. 
\labl{anomOne}
}
(In principle this gauge transformation could also contain the 
contribution $\sum_I \gx_I  \d \ga_0 \ge(y -I)$ with $\ge(y)$ the
appropriate step function. But by a suitable chosen
compensating gauge transformation \eqref{gaugecA} this transformation
can be turned into the one given above). Note that this anomalous
gauge transformation only affects 
$\gd_{\ga_0}\cA_5 = \sum_I \gx_I \ga_0\gd(y - I)$, while the four dimensional
components are invariant $\gd_{\ga_0} \cA_m = 0$. 
After substituting (\ref{B2eom}) into the action (\ref{GSdual}) the one--form 
formulation of the Green--Schwarz mechanism becomes
\equa{
S_{GS1} & = \int \! \Big[ 
\frac 12 *\! \d A_1 \d A_1 + 
\frac 12 \! * \!  \big( \d \cA_1 \!+\! * \go_3(A) \!-\! X_2(A)\big) 
 \big( \d \cA_1 \!+\! * \go_3(A) \!-\! X_2(A)\big) 
+ \frac 12 * \go_3(A) \go_3(A)
\Big] 
\non \\[1ex] 
& = \int \Big[ 
\frac 12 *\! \d A_1 \d A_1 + 
\frac 12 *\! \d \cA_1 \d \cA_1 
- \go_3(A) \d \cA_1 
- \Big(  * \d \cA_1 - \frac 12 *\! X_2(A) \Big) X_2(A) 
\Big]~. 
\labl{Vaction}
}
Notice that the term $\go_3(A) X_2(A)$ is zero. If there are
multiple non-anomalous $\U{1}$'s then this term does not cancel 
and leads to the generalized Chern--Simons terms with two gauge field 
one--forms~\cite{afl}. The supersymmetrization of this possibility is
described in Appendix~\ref{sc:MultiU1}.

The term $- \go_3(A) \d \cA_1$ is not gauge
invariant, and therefore the gauge variation of \eqref{Vaction} does
not equal \eqref{varGS}, unless the four dimensional components of the
Green--Schwarz vector vanishes at the boundaries 
$\cA_\gm|_I = 0$. This action is only defined formally since it
involves squares of delta functions:  
\equ{
- \int  \Big(* \d \cA_1 - \frac 12 *\! X_2(A) \Big)X_2(A)  
= \frac 12 \sum_I \gx_I \int \d^5 x 
\Big( A^m \cF_{m5} - \frac 12  \gx_I \gd(0) A_m A^m \Big)\big|_I
~. 
\labl{Singular}
}
However, since it has been obtained from the regular action \eqref{GSdual},
there is a unique and well--defined way to deal with the  singular
contribution $\gd(0)$ in this case.

The dual formulation using the two--form $\cB_2$
with $\smash{\widehat \cF_2} = -* \d \cB_2$ is manifestly
free of any singular looking terms 
\equ{
S_{GS2} = 
\int \Big[\frac 12 * \d A_1 \d A_1 
+   \frac 12 * \! \big( \d \cB_2 + \go_3(A) \big)
\big( \d \cB_2 + \go_3(A) \big) 
- \d \cB_2 X_2(A) \Big]~.  
\labl{GS2action}
}
This is the five dimensional analog of the ten dimensional action of
the anti--symmetric tensor in the supergravity multiplet coupled to
the super Yang--Mills theory which is, for example, relevant for the
low energy description of the heterotic string on orbifolds. In this case
one similarly finds that the two--form description is free of singular
terms, while in the dual formulation involving a six--form the
singular $\gd(0)$ term arises as well, as can be inferred from ref.\ 
\cite{GrootNibbelink:2003gb}.

Furthermore, by introducing the gauge
invariant three--form field strength $\widehat \cH = \d \cB_2 + \go_3(A)$ 
one obtains the anomalous Bianchi identity $\d \widehat \cH = F_2^2$.
Finally, the duality relation \eqref{VTduality} between the two-- and
one--form formulation is modified to 
\equ{
-* \d \cB_2 = \widehat \cF_2 = \d \cA_1 + *\go_3(A) - X_2(A)~. 
\labl{onetwodual}
}

\subsection{The four dimensional anomalous U(1) mass spectrum}
\labl{sc:U1mass}

We have seen that there are two equivalent dual formulations
of the Green--Schwarz mechanism in five dimensions. Let us now
determine the four dimensional mass spectrum of the anomalous
$\U{1}_A$ photon field together with the Green--Schwarz vector or
tensor fields. At first sight it seems that the two formulations lead
to very different mass spectra, since in the vector formulation
(\ref{Singular}) there are divergent boundary mass terms which are 
absent in the dual tensor formulation (\ref{GS2action}). This would then
contradict the equivalence of the two dual formulations. However 
we will see that both formulations give rise to the same mass spectrum
in four dimensions. The key observation is that both the vector and
tensor formulations contain mixing terms between the vector or tensor field
and the anomalous $\U{1}_A$ gauge fields that need to be correctly
taken into account. This is done by showing that both formulations 
lead to the same recursion relations for the components of the 
eigenvectors, that will enable us to determine the mass spectrum explicitly. 
Since the method of computing with infinite dimensional mass matrices 
is well--known~\cite{ddg1}, only the essential steps required for
demonstrating how the duality is manifesting itself will be shown.

Since theories of five dimensional vector fields with boundary masses are
more familiar, we consider the vector formulation first. We only require
the quadratic part of \eqref{Vaction} and use the gauge fixing conditions 
\equ{
\der^m A_m + \der_5 A_5 = 0~, 
\qquad 
\der^m \cA_m + \der_5 \cA_5 = 0~, 
}
to show that the $A_5$ and $\cA_5$ are decoupled from the gauge
fields $A_m$ and $\cA_m$. Using the standard Kaluza--Klein
decompositions on the orbifold
\equ{
 A_m = \sum_{n \geq 0} \get_n A_m^{(n)}\cos \Big( \frac {ny}R \Big)~, 
\qquad 
 \cA_m = \sum_{n \geq 1} \get_n \cA_m^{(n)}\sin \Big( \frac {ny}R \Big)~,  
}
with $\get_0^2 = \frac 1{\gp R}$ and 
$\get_{n\neq 0}^2  = \frac 2{\gp R}$, 
we find the following recursion relations for the Kaluza-Klein modes 
\equa{ 
n \geq 0:~~
\Big( m^2 - \big( \frac {n}{R} \big)^2 \Big)  A^{(n)}_m 
= & \
\frac 12 \sum_{n' \geq 1} \get_n \get_{n'} \frac {n'}R 
\Big( \gx_0 + \gx_\gp (-1)^{n+n'} \Big)  \cA^{(n')}_m 
\non \\[1ex] &\  
+ \frac 12 \gd(0) \sum_{n''\geq 0} \get_n \get_{n''} 
\Big( \gx_0^2 + \gx_\gp^2 (-1)^{n+n''} \Big)  A^{(n'')}_m~, 
\\[1ex] 
n' \geq 1:~~
\Big( m^2 - \big( \frac {n'}{R} \big)^2 \Big)  \cA^{(n')}_m
=  & \ 
\frac 12 \sum_{n'' \geq 0} \get_{n'} \get_{n''} \frac {n'}R 
\Big( \gx_0 + \gx_\gp (-1)^{n'+n''} \Big)  A^{(n'')}_m~.
}
In the above expressions we have set the four dimensional momentum 
squared $\der_4^2$ equal to the mass eigenvalue $m^2$, and we have 
explicitly shown the dependence on the orbifold radius $R$. Note that
$ \cA_m^{(n')}$ can be solved from the second equation and substituted 
into the first relation, where the resulting infinite divergent sums and 
the $\gd(0)$ are regularised in the following way~\cite{mp}
\equ{
\gd(0) = \frac 1{2\gp R} 
\Bigg[
\frac {m^2}{m^2} 
+ 2 \sum_{n' \geq 1} \frac {m^2 - (n'/R)^2}{m^2 - (n'/R)^2}
\Bigg]~, 
\qquad 
\frac 2{\gp R} \sum_{n' \geq 1} (-1)^{n'} = - \frac 1{\gp R}~.
\labl{RegDelta}
} 
This gives rise to an equation solely for the modes $A_m^{(n)}$,
which has the form 
\equa{
\Big( m^2 - \big( \frac{n}{R} \big)^2 \Big)  A^{(n)}_m = 
\frac 14 \sum_{n',n'' \geq 0} \get_n \get_{n''} 
\Bigg[ & 
\big( \gx_0^2 + \gx_\gp^2 (-1)^{n + n''} \big) 
\frac {\get_{n'}^2}{1 - \big(\frac {n'}{mR}\big)^2} 
\non \\[1ex] & \ 
+ \gx_0 \gx_\gp \big( (-1)^n + (-1)^{n''} \big) 
\frac {\get_{n'}^2(-1)^{n'}}{1 - \big(\frac {n'}{mR}\big)^2} 
\Bigg]  A^{(n'')}_{m}~. 
\labl{EVrec} 
}
This equation encodes the mass eigenvalues of the anomalous $\U{1}_A$ photon 
resulting from the Green--Schwarz theory coupled to the anomalous 
$\U{1}_A$ gauge theory. After some algebra one finds the mass eigenvalue 
equation and solution
\equ{
\tan^2 (\gp m R)  = 
\frac 14  
\Big( \frac {\gx_0 + \gx_\gp}{ 1 - \frac 14 \gx_0 \gx_\gp} \Big)^2 
\quad \Ra \quad 
m = \frac n R + \frac 1{\pi R} \arctan \Big( 
\frac 12 \frac {\gx_0 + \gx_\pi}{1 - \frac 14 \gx_0 \gx_\pi} 
\Big)~. 
\labl{u1Amass}
} 
We see that the anomalous photon mass is finite at tree level, and the only
scale it depends on is the orbifold radius $R$. 
In the limit that $\xi_0+\xi_\pi \ll 1$ the 
complete Kaluza--Klein tower shifts approximately by the amount
\(
\gd m \approx |\gx_0 + \gx_\pi|/(2\pi R).
\)
This result is compatible with the shift of the zero mode of the anomalous
$\U{1}_A$ gauge field in four dimensions derived in ref.\ \cite{akr}. 
Notice also that when $\xi_0+\xi_\pi = 0$ the Kaluza--Klein spectrum 
(\ref{u1Amass}) remains unchanged, in particular the zero mode photon 
stays massless.

To check that the dual formulations are equivalent let us next 
explain how the same recursion relation \eqref{EVrec} is
obtained in the tensor formulation. Following the same methodology as
in the vector formulation, we will consider the quadratic part of
\eqref{GS2action} and use the gauge fixing conditions  
\equ{
\der^m A_m + \der_5 A_5 = 0~, 
\qquad 
\der^m \cB_{mn} + \der_5 \cB_{5n} = 0~, 
}
to decouple $A_5$ and $\cB_{5n}$ from $A_m$ and $\cB_{mn}$, respectively. 
In terms of the Kaluza-Klein mode expansions
\equ{
A_m = \sum_{n\geq 0} \get_n A_m^{(n)} \cos \big( \frac {ny}R \big)~, 
\qquad 
\cB_{mn} = \sum_{n\geq 0} \get_n \cB_{mn}^{(n)} \cos \big( \frac {ny}R \big)~, 
}  
the recursion relations for $n,n' \geq 0$ are given by 
\equa{
\Big( \der_4^2 - \big( \frac{n'}{R} \big)^2 \Big) \cB_{st}^{(n')} 
= & \
- \frac 12 \ge_{st}{}^{pq} \sum_{n''\geq 0} \get_{n'} \get_{n''} 
\big( \gx_0 + \gx_\gp (-1)^{n' + n''} \big) \der_p A_q^{(n'')}~,
\\[1ex]
\Big( \der_4^2 - \big( \frac{n}{R} \big)^2 \Big) A_{m}^{(n)}
= & \
-\frac 14 \ge_{m}{}^{rst} \sum_{n'\geq 0} \get_{n} \get_{n'} 
\big( \gx_0 + \gx_\gp (-1)^{n + n'} \big) \der_r \cB_{st}^{(n')}~.
}
Note that we cannot replace $\der_4^2$ by the
mass eigenvalues in the above equations  since there are single derivatives 
in these recursion relations. Instead to eliminate the single derivatives 
we solve the first equation for $\cB_{st}^{(n')}$ and then substitute 
this into the second equation to obtain 
\equa{
\Big( \der_4^2 - \big( \frac{n}{R} \big)^2 \Big) A_{m}^{(n)} 
= 
\frac 14 \sum_{n',n''\geq 0} & \get_n \get_{n''} \get_{n'}^2 
\Big( \gx_0^2 + \gx_\gp^2 (-1)^{n+n''} 
\non \\[1ex] & \ 
+ (-1)^{n'} \gx_0 \gx_\gp  \big( (-1)^n + (-1)^{n''} \big) 
\Big)  
\frac { \gd_m^q \der_4^2 - \der_m \der^q}
{\der_4^2 - (n'/R)^2} A^{(n'')}_q~. 
} 
Notice that the last term contains the projector onto the transverse
part of the gauge field. Therefore, for the transverse polarized
gauge fields, we can replace $\der_4^2$ with the mass eigenvalues $m$,
and obtain an equation identical to \eqref{EVrec}. Consequently, 
the anomalous $\U{1}_A$ mass spectrum in the vector and 
tensor formulation is the same, and the two dual formulations are indeed
equivalent. However, the interesting aspect of the tensor formulation is that 
we never had to resort to the --in principle-- ill--defined regularization 
of $\gd(0)$ given in \eqref{RegDelta}. Thus, the agreement of the Kaluza-Klein
mass spectrum in the vector and tensor formulation
justifies this regularization prescription.

\section{Supersymmetric vector/tensor multiplet duality}
\labl{sc:SusyVT}

Before describing the full supersymmetric Green--Schwarz
mechanism in five dimensions, we need to introduce the five
dimensional vector and tensor multiplets. The five dimensional vector 
multiplet has been discussed in various places using four dimensional
$N\!=\!1$ superfields in the literature
\cite{Marcus:1983wb,Arkani-Hamed:2001tb,Hebecker:2001ke}, while the
five dimensional tensor multiplet is less known in this formalism. 
For completeness and to fix our notation we describe both these
multiplets in the following two subsections. We also give the straightforward
generalization to multiple vector multiplets which will proof
convenient in later sections. We will use the conventions from
Wess and Bagger~\cite{Wess:1992cp} throughout this paper. 
For simplicity the component forms of the supermultiplets and actions 
will be restricted to bosons only. Since all expressions will be
given in four dimensional $N\!=\!1$ superspace, it is straightforward,
though tedious, to obtain the fermionic terms. In this section
we first focus on the vector multiplet that occurs in the supersymmetric 
Green--Schwarz mechanism, and a collection of $\U{1}$ vector multiplets in
general. After that we describe the five dimensional tensor multiplet
and discuss the supersymmetric extension of the duality between
vector and tensor fields.  (The anomalous $\U{1}_A$ vector multiplet will be
discussed in section \ref{sc:SusyGS} together with a description of the 
supersymmetric version of the Chern--Simons three--form.)

\subsection{Five dimensional vector multiplet}
\labl{sc:VecMplt}

A vector multiplet in five dimensions contains a vector field $\cA_M$,
a real scalar $\gvf$ and a Dirac (or symplectic Majorana) fermion. By
reducing to four dimensional supersymmetry one can infer that the
five dimensional vector multiplet is described by a vector superfield
$\cV^\dag = \cV$  and a chiral superfield $\cS$ with 
$\bD_\dga \cS = 0$. The super covariant derivatives of the four dimensional 
$N\!=\!1$ superspace are denoted by $D_\ga$ and $\bD_\dga$. 
The gauge transformation \eqref{gaugecA} of $\cA_M$ is lifted in
$N\!=\!1$ superspace to
\equ{
\gd_{\gPs} \cV = \gPs + \bgPs~, 
\qquad 
\gd_{\gPs} \cS = \sqrt 2 \, \der_5 \gPs~,
}
where $\gPs$ is a chiral superfield. (The gauge transformation 
\eqref{gaugecA} is obtained by taking $\gPs = i \gb_0$.) 
The bosonic components of these superfields for the vector multiplet
$\cV$ in the Wess--Zumino gauge and the chiral multiplet $\cS$ are  
identified by   
\equ{
- \frac 12 [ D_\ga, \bD_\dga ] \cV| = \gs^m_{\ga \dga} \, \cA_m~, 
\quad   
\frac 18 D^\ga \bD^2 D_\ga \cV| = \cD~,
\quad  
\cS | = \frac 1{\sqrt 2} \big( \gvf + i \cA_5 \big)~, 
\quad 
- \frac 14 D^2 \cS | =  \cF~,  
}
where we have introduced the four dimensional indices 
$m = 0, \ldots 3$. As usual the notation $|$ indicates that
we have set all $\bgth_\dga = \gth_\ga = 0$. Out of the superfields $\cV$ and
$\cS$ two independent super gauge invariant superfields can be
constructed  
\equ{
\cW_\ga = - \frac 14 \bD^2 D_\ga \cV~, 
\qquad 
\cV_5 =\frac 1{\sqrt 2} ( \cS + \bar \cS) -  \der_5 \cV~.  
\labl{superFieldStrength}
}
The bosonic components of these superfields are given by
\equ{
\arry{c}{ \dsp 
D_{\gb} \cW_{\ga}| = 
- i (\gs^{mn} \ge)_{\gb\ga} \cF_{mn} - \ge_{\gb\ga} \cD~, 
\qquad 
\cV_5| = \gvf~, 
\qquad 
- D^2 \cV_5| = 2 \sqrt 2\, \cF~, 
\\[1ex] \dsp 
- \frac 12 [D_\ga, \bD_\dga] \cV_5| =  \gs^m_{\ga \dga} \cF_{m5}~, 
\qquad 
\frac 1{8} D^\ga \bD^2 D_\ga \cV_5| = - \der_5 \cD~,
}
} 
where $\cF_{mn} = \der_m \cA_n - \der_n \cA_m$ and 
$\cF_{m 5} = \der_m \cA_5 - \der_5 \cA_m$. The action for the five
dimensional super Yang--Mills theory can be represented as 
\cite{Arkani-Hamed:2001tb} 
\equa{
S_{SV} &= \int \d^5 x \left[
 \int \d^2 \gth \frac 14 \cW^\ga(\cV) \cW_\ga(\cV) + \text{h.c.}
 + 
\int \d^4 \gth \cV_5^2 
\right]
\labl{AcSV} 
\\[1ex] 
&\supset 
\int \d^5 x \Big[ 
- \frac 14 \cF_{MN} \cF^{MN} - \frac 12 \der_M \gvf \der^M \gvf 
+ \frac 12 \big( \cD + \der_5 \gvf \big)^2 + \bcF \cF 
\Big]~. 
\labl{AcSVcomp} 
}
The second line is obtained by restricting to the bosonic part of the
action; we denote this by $\supset$. 
We have completed the square involving the auxiliary field $\cD$
to show that up to the auxiliary field equation of motion the
component action is Lorentz invariant in five dimensions, because in
later section we encounter terms that will modify the auxiliary field
equations.

The present discussion can easily be extended to the case of multiple
five dimensional $\U{1}$ vector multiplets described by the
superfields $(\cV^a, \cS^a)$ and labeled with lowercase Latin indices 
$a,b, \ldots$ run over all the vector multiplets, anomalous and
non--anomalous. 
The most general gauge invariant action is encoded by a prepotential
$\cP(\gvf)$. This real function of the real scalars $\gvf =(\gvf^a)$ is at
most a cubic polynomial~\cite{Seiberg:1996bd}
\equ{
\cP(\gvf) = \frac {1}{2} c_{ab} \gvf^a \gvf^b 
+ \frac {1}{6} c_{abc} \gvf^a \gvf^b \gvf^c~, 
\label{RealPre}  
}
where the coefficients $c_{ab}$ and $c_{abc}$ are real and totally
symmetric. Denoting differentiation w.r.t.\ $\gvf^a$ by 
$\cP_a \equiv \partial\cP/\partial\gvf^a$, the superspace
expression for the general action with an arbitrary prepotential $\cP$ reads  
\equa{
S_{\cV}  & = \! \int \!\!  \d^5 x 
\bigg[ 
\int \!\! \d^2 \gth \frac 14 \Big( 
\cP_{ab}(\sqrt 2 \cS) \cW^{a\ga} \cW^b_\ga 
- \frac 1{6} \cP_{abc} 
\bD^2 \big( \cV^a D^\ga \der_5 \cV^b \!-\! D^\ga \cV^a \der_5 \cV^b \big)
\cW^c_\ga  
\Big)
+ \text{h.c.} 
\non \\[1ex] & 
\qquad \qquad + \int \!\!\d^4 \gth  \,  
2\cP(\cV_5) 
\bigg]
\labl{AcPreV} \\[1ex]
& \supset  \!\int\!\! \d^5 x 
\Big[ 
\cP_{ab}(\gvf) 
\Big(  
- \frac 14 \cF^a_{MN} \cF^{b MN}
 - \frac 12 \der_M \gvf^a \der^M \gvf^b 
+ \frac 12 (\cD + \der_5 \gvf)^a (\cD + \der_5 \gvf)^b 
+ \bcF^a \cF^b 
\Big) 
\non \\[1ex]
&\qquad\qquad
+ \frac 1{24} \cP_{abc} \ge^{MNPQR}  \cA_M^a \cF^b_{NP} \cF^c_{QR} 
\Big]~.
\labl{AcPreVcomp}
} 
In order for the general action to reproduce the kinetic action for 
a single vector multiplet \eqref{AcSV} the normalization of the 
prepotential is chosen such that $\cP_{SV} = \frac 12 \gvf^2$.
The gauge variation of \eqref{AcPreV} is given by 
\equ{
\gd_\gPs S_{\cV} = \! \int \!\!  \d^5 x \bigg(
\int \d^2 \gth \ 
\frac 13 \cP_{abc} \gPs^a \cW^{b\ga} \cW^c_\ga
+ \text{h.c.} 
\bigg)\big( \gd(y - \gp) - \gd(y) \big)~. \labl{var} 
}


\subsection{Five dimensional tensor multiplet}
\labl{sc:TenMplt}

A tensor multiplet contains an
anti--symmetric rank two tensor $\cB_{MN}$, a real scalar $\gs$ and a
Dirac fermion. Using the four dimensional superspace language these
states can be described by a chiral spinor multiplet $\cT_\ga$, with
$\bD_\dga \cT_\ga =0$, and a vector multiplet $\cU$ with 
$\cU^\dag = \cU$. The 
gauge transformation \eqref{TwoFormGauge} of the two--form $\cB_2$ 
contains both a four dimensional vector $\gb_m$ and a scalar $\gb_5$,
$\gb_1 = \gb_m \d x^m + \gb_5 \d x^5$. 
The extension to a superspace transformation is
\equ{
\gd \cT_\ga = - \frac 14 \bD^2 D_\ga \cC~, 
\qquad 
\gd \bcT_\dga = - \frac 14 D^2 \bD_\dga \cC~, 
\qquad 
\gd \cU = \gPs + \bgPs - \der_5 \cC~,  
}
where $\cC$ and $\gPs$ denote a vector and chiral superfield, respectively. 
(The four dimensional analog of this transformation is well--known, 
see for example \cite{Gates:ay,Gates:1983nr}.) This means that one can define 
a Wess--Zumino gauge for both $\cU$ and $\cT_\ga$. For the chiral superfield
$\cT_\ga$ this gauge is defined such that the $\ge_{\gb\ga}$ part of
$D_\gb \cT_\ga$ is purely imaginary. In the Wess--Zumino gauge the bosonic 
components of the supermultiplets $\cT_\ga$ and $\cU$ are given by
\equ{
D_{\gb} \cT_{\ga}| = 
- i (\gs^{mn}\ge)_{\gb\ga} \cB_{mn} - \ge_{\gb\ga} i \gs~,  
\quad 
-\frac 12 [D_\ga, \bD_\dga] \cU| = \gs^m_{\ga \dga} \cB_{m5}~, 
\quad 
\frac 18 D^\ga \bD^2 D_\ga \cU| = \cD_T~. 
} 
Furthermore two gauge invariant superfields can be 
constructed from $\cT_\ga$ and $\cU$: 
\equ{
\cT_{\!5\,\ga} =  \der_5 \cT_\ga + \cW_\ga(\cU)~,
\qquad 
\cL = \frac i4  (D^\ga\cT_\ga - \bD_\dga \bcT^\dga)~. 
}
The supermultiplet $\cT_{\!5\, \ga}$ is chiral, 
$\bD_\dga \cT_{\!5\, \ga} = 0$, while $\cL$ defines a linear multiplet  
$D^2 \cL = \bD^2 \cL =~0$~\cite{Binetruy:1991sz,Girardi:1998ju}. The 
bosonic components of these gauge invariant superfields read  
\equ{
\arry{c}{\dsp 
D_\gb \cT_{\!5\,\ga}| = 
- i (\gs^{mn} \ge)_{\gb\ga} \cH_{mn5}  
- \ge_{\gb\ga} \big(\cD_T +i \der_5 \gs \big)~, 
\\[1ex] \dsp 
- \frac 12 [ D_\ga, \bD_\dga] \cL|  = - \frac 16 
\ge^{mnkl} \cH_{mnk} (\gs_l)_{\ga\dga}, 
\qquad 
\cL | = \gs~, 
}
}
where
$\cH_{mnk} = \der_m B_{nk} + \der_n B_{km} + \der_k B_{mn}$ and 
$\cH_{mn5} = \der_m B_{n5} - \der_n B_{m5} + \der_5 B_{mn}$. 
Hence the action for the five dimensional tensor multiplet can 
be represented as 
\equa{
S_{ST} & = 
\int \d^5 x \bigg[ 
\int \d^2 \gth 
\frac 14 \cT_{\!5}^\ga \cT_{\!5\, \ga} + \text{h.c.} + 
\int \d^4 \gth (-\! \cL^2) 
\bigg] 
\labl{AcST} 
\\[1ex] 
& \supset  \int \d^5 x 
\Big[
-\! \frac 1{12} \cH_{MNK} \cH^{MNK}  - \frac 12 \der_M \gs \der^M \gs 
+ \frac 12 \cD_T^2 
\Big]~. 
\labl{AcSTcomp}
}
The second line only contains the bosonic part of the action.
Unlike the vector multiplet action the tensor superfield action is manifestly
five dimensional Lorentz invariant off--shell, and the equation of motion
of $\cD_T$ is trivial.


\subsection{The supersymmetric vector/tensor duality} 
\labl{sc:SusyDual}

In the previous two subsections we have discussed the five dimensional
vector and tensor multiplet using $N\!=\!1$ four dimensional
superspace techniques. We now wish to supersymmetrize the five
dimensional vector/tensor duality described in section
\ref{sc:VecTenDual}. Proceeding as in that section we will start with
a hybrid description in which either the vector or tensor formulation has not yet
been chosen. It is useful to rewrite \eqref{FormDual5D} as  
\equa{
S_{TV} =  \int \d^5 x \Big[ &  
- \frac 14 \widehat \cF_{mn} \widehat \cF^{mn} 
- \frac 12 \widehat \cF_{m5} \widehat \cF^{m5} 
- \frac 14 \ge^{mnkl} \cH_{mn5} \widehat \cF_{kl}  
+ \frac 14 \ge^{mnkl} \cB_{mn} (\der_k \widehat\cF_{l5} - \der_l \widehat\cF_{k5}) 
\Big]~.
\labl{AcDualPre}
}
Therefore, to obtain the supersymmetric version we see that 
apart from the superfields $\cT_\ga$ and $\cU$ describing
the tensor multiplet degrees of freedom, we also need to introduce a chiral
superfield, $\smash{\widehat \cW_\ga}$, and a vector superfield,
$\smash{\widehat \cV_5}$, for the non--dynamical (hatted) fields
$\smash{\widehat \cF_{mn}}$ and $\smash{\widehat \cF_{m5}}$.  
Since only the field strength of the vector $\smash{\widehat\cF_{m5}}$ 
appears in \eqref{AcDualPre},  the supersymmetric generalization will
involve only the superfield strength
$\cW_\ga(\smash{\widehat\cV_5})$. Using standard superfield
methods to obtain component actions we find that 
the supersymmetric generalization of \eqref{AcDualPre} is a part of
\equ{
S_{STV} = \int \d^5 x \bigg[ 
\int \d^2 \gth \Big( 
\frac 14 \widehat \cW^\ga \widehat \cW_\ga 
- \frac i2 \cT_5^\ga \widehat \cW_\ga 
+ \frac i2 \cT^\ga \cW_\ga(\widehat \cV_5) 
\Big) + \text{h.c.} + \int \d^4 \gth\,  \widehat \cV_5^2 \bigg] 
\supset S_{TV}~.
\labl{AcDualSusy}
}
Note that the last term of $S_{STV}$ 
is not present in \eqref{AcDualPre}, but is needed to
ensure that the whole action is five dimensional Lorentz invariant.

The supersymmetric realization of the vector/tensor duality can now be
described as follows (the related scalar/tensor duality for $N\!=\!2$ in
four dimensions has been discussed in \cite{Lindstrom:1983rt}): 
The supersymmetric version of the elimination of $\cB_{MN}$ is
implemented by varying with respect to the corresponding superfields
$\cU$ and $\cT_\ga$. 
Because of the chirality of $\smash{\widehat\cW_\ga}$ the equation of motion of
$\cU$ implies that  $\smash{\widehat\cW_\ga}$ can be expressed in terms of a
vector superfield $\cV$:   
\equ{
\bD_\dga \overline{\widehat\cW}{}^\dga - D^\ga \widehat \cW_\ga = 0
\qquad \Ra \qquad 
\widehat \cW_\ga = \cW_\ga(\cV) = - \frac 14 \bD^2 D_\ga \cV~. 
\labl{solcW}
}
To work out the consequences of the
variation of $\cT_\ga$, it proves useful to write this superfield as 
$\cT_\ga = \bD^2 \cC_\ga$, where $\cC_\ga$ is an unconstrained spinor 
multiplet. Using the solution \eqref{solcW} for $\smash{\widehat\cW_\ga}$, we 
obtain constraints which are readily solved by introducing a chiral superfield
$\cS$ 
\equ{
\bD^2 D_\ga ( \widehat\cV_5 + \der_5 \cV) = 
D^2 \bD_\dga ( \widehat\cV_5 + \der_5 \cV) = 0
\quad \Ra \quad
\widehat \cV_5 = \cV_5 
= \frac 1{\sqrt 2} ( \cS + \bar \cS) -  \der_5 \cV~. 
\labl{solcV5}
}
Substituting the solutions \eqref{solcW} and \eqref{solcV5} 
for $\smash{\widehat \cW_\ga}$ and $\smash{\widehat \cV_5}$,
respectively, back into the supersymmetric action \eqref{AcDualSusy}
leads to  the action $S_{SV}$ of a vector multiplet, given in \eqref{AcSV}.

Alternatively, to eliminate the supersymmetric extension of $\cF_{MN}$, i.e.\ 
$\smash{\widehat \cW_\ga}$ and $\smash{\widehat \cV_5}$, we rewrite the action
\eqref{AcDualSusy} in the form 
\equ{
S_{STV} = \int \d^5 x \bigg[ 
\int \d^2 \gth \Big(
\frac 14 \widehat \cW^\ga \widehat \cW_\ga 
- \frac i2 \cT_5^\ga \widehat \cW_\ga
\Big) + \text{h.c.} + 
\int \d^4 \gth \Big( \widehat \cV_5^2 
- \frac i2  (D^\ga \cT_\ga - \bD_\dga \bcT^\dga) \widehat \cV_5 \Big)
\bigg]~,
\labl{actionSTV}
} 
where we have used the chirality of $\cT_\ga$ and 
$\cW_\ga(\smash{\widehat \cV_5})$, and replaced $\int d^2 \gth$ by $-\frac 14 D^2$ 
under the spacetime integral. The resulting equations of motion
for the non--dynamical superfields $\smash{\widehat \cW_\ga}$ and
$\smash{\widehat \cV_5}$ are  
\equ{
\widehat \cW^\ga = i \cT_{\!5\, \ga}~,
\qquad 
\widehat \cV_5 = \cL = \frac i4  (D^\ga\cT_\ga - \bD_\dga \bcT^\dga)~.
\labl{solcWcV5}
}
Substituting these expressions back into (\ref{actionSTV}) gives
the action $S_{ST}$ of a tensor multiplet, see \eqref{AcST}.

By combining both expressions for $\widehat \cV_5$ and $\widehat \cW_\ga$ 
the supersymmetric version of the duality relation \eqref{onetwodual} is
found to be
\equ{
\frac 1{\sqrt 2} (\cS + \bar \cS) - \der_5 \cV  
= \widehat \cV_5 = \frac i4  (D^\ga \cT_\ga - \bD_\dga \bcT^\dga)~, 
\qquad 
\cW_\ga(\cV) = \widehat \cW_\ga = i ( \der_5 \cT_\ga + \cW_\ga(\cU))~.
\labl{AcPrePot}
}
It is not surprising that the five dimensional duality is written in terms of 
two equations since one also obtains two equations if the duality 
relation \eqref{onetwodual} is dimensionally reduced and written in
four dimensional component language. Notice also that the lowest
component of the first equation in (\ref{AcPrePot})
gives a linear version of the duality relation in \cite{Seiberg:1996bd},
namely
\equ{
\gs  = \gvf = \frac{\der \cP_{SV}}{\der \gvf}~. 
}

\section{Supersymmetric Green--Schwarz mechanism}
\labl{sc:SusyGS}

After presenting the duality between the vector and tensor multiplets
in five dimensions, we now proceed to give an $N\!=\!1$ four dimensional
superspace 
discussion of the Green--Schwarz mechanism in five dimensions. We
begin by introducing the Chern--Simons superfields for the anomalous
vector multiplet. This will enable us to describe the
supersymmetric version of the Green--Schwarz theory in the vector and
tensor multiplet representation. 


\subsection{Chern--Simons three--form superfields} 
\labl{sc:AnomVecMplt} 

The anomalous vector multiplet in five dimensions is defined similarly
to the non--anomalous vector multiplet in section \ref{sc:VecMplt}, 
except that we use standard italic letters for the fields instead of 
calligraphic letters to distinguish the two. For notational
completeness we begin by repeating the relevant definition of the
vector multiplet in five dimensions so that the subsequent discussion
on the supersymmetrization of the Chern--Simons three--form in five
dimensions will be self contained. We introduce a vector
multiplet $V$ and a chiral multiplet $S$ with bosonic components 
\equ{
- \frac 12 [ D_\ga, \bD_\dga ] V| = \gs^m_{\ga \dga} \, A_m~, 
\quad   
\frac 18 D^\ga \bD^2 D_\ga V| = D~,
\quad  
S | = \frac 1{\sqrt 2} \big( \gf + i A_5 \big)~, 
\quad 
-\frac 14 D^2 S | =  F~,  
}
where $V$ is in Wess--Zumino gauge. The bosonic components of 
the super gauge invariant superfields 
\equ{
W_\ga = - \frac 14 \bD^2 D_\ga V~, 
\qquad 
V_5 =\frac 1{\sqrt 2} ( S + \bar S) -  \der_5 V~,
\labl{WV5defns}
}
are given by
\equ{
\arry{c}{ \dsp 
D_{\gb} W_{\ga}| = 
- i (\gs^{mn} \ge)_{\gb\ga} F_{mn} - \ge_{\gb\ga} D~, 
\qquad 
V_5| = \gf~, 
\qquad 
- D^2 V_5| = 2 \sqrt 2\, F~, 
\\[1ex]  \dsp 
- \frac 12 [D_\ga, \bD_\dga] V_5| =  \gs^m_{\ga \dga} F_{m5}~, 
\qquad  \dsp 
\frac 1{8} D^\ga \bD^2 D_\ga V_5| = - \der_5 D~. 
}
}

With these definitions we can now discuss a supersymmetrization of the 
Chern--Simons three--form in five dimensions. The (4+1) dimensional
decomposition of the definition \eqref{CS3form} of this three--form
\equ{
\frac 16 \der_m \go_{npq} \ge^{mnpq} 
= \frac 14 F_{mn} F_{pq} \ge^{mnpq}~, 
\quad 
\frac 12 \der_m \go_{np5}  \ge^{mnpq} = 
\Big(
F_{mn} F_{p5} - \frac 16 \der_5 \go_{mnp} 
\Big) \ge^{mnpq}~,
\labl{dimredCS3}
}
implies that we need to introduce two Chern--Simons superfields in five
dimensions in order to contain all the degrees of freedom. 
In appendix \ref{sc:SCSdefs} we describe a convenient way
to determine the supersymmetrizations of the definitions. 
For the first relation in (\ref{dimredCS3}) we use the four dimensional
Chern--Simons superfield $\gO$ which is a real vector
superfield defined by the equations  
\cite{Cecotti:1987nw,Bellucci:1988ff,Binetruy:1991sz,Derendinger:1994gx,Girardi:1998ju} 
\equ{
\bD^2 \gO = W^\ga W_\ga~, 
\qquad 
D^2 \gO = \bW_\dga \bW^\dga~. 
\labl{defCSsuper}
}
The solution to these equations is $\gO = \gO(V,V)$ where we have
defined 
\equ{
\gO(V,V') = - \frac 14 \big( 
D^\ga V W_\ga(V') + \bD_\dga V \bW^\dga(V') +  V D^\ga W_\ga(V')  
\big)~. 
\labl{gOgen}
} 
Since $D^\ga W_\ga = \bD_\dga \bW^\dga$ the last term in this
expression may also be written as $V  \bD_\dga \bW^\dga$. 
In Wess--Zumino gauge the four dimensional components 
$\go_{mnp} = A_m F_{np} + A_n F_{pm} + A_p F_{mn}$ 
of the Chern--Simons three--form \eqref{CS3form} are obtained by the
restriction  
\equ{
-\frac 12 [ D_\ga, \bD_\dga] \gO | \supset 
 \frac 1{12} \go_{mnk} \ge^{mnkl} (\gs_l)_{\ga \dga}~. 
}

The supermultiplet containing the additional components  
$\go_{mn5} = A_m F_{n5} - A_n F_{m5} + A_5 F_{mn}$ 
of the five dimensional Chern--Simons three--form $\go_3$ (see the 
second relation in (\ref{dimredCS3}) and Appendix \ref{sc:SCSdefs}),
is obtained by solving the equation 
\equ{
\frac i 2 \big( D^\ga Y_\ga - \bD_\dga \bY^\dga \big) 
= - 4 \der_5 \gO(V, V) - 8 \gO(V_5, V)~, 
\labl{defCSsuper5D}
}
where $\gO(V,V')$ is defined in \eqref{gOgen}. 
The solution is described by the chiral spinor multiplet $Y_\ga$ and its
conjugate, where
\equ{
Y_\ga = \frac i4 \bD^2 \big(
V_5 D_\ga V 
- D_\ga V_5  V 
+ \sqrt 2 \, S D_\ga V 
\big)~, 
\qquad 
\bD_\dga Y_\ga = 0~.
\labl{solY}
}
(To prove that (\ref{solY}) is indeed a solution of the
superfield equation (\ref{defCSsuper5D}), it is useful to
employ  the identities \eqref{idenVV} and \eqref{der5gO} of 
Appendix \ref{sc:SCSdefs}). 
After dropping all fermionic
terms in the Wess--Zumino gauge the component $\go_{mn5}$ appears in the
projection
\equ{
D_{\gb}  Y_{\ga}| \supset - i (\gs^{mn}\ge)_{\gb\ga} 
\big( 
\go_{mn5} -2 i \gf F_{mn} 
\big) 
- \ge_{\gb\ga} 
\big( 
A_5 D - \der_m \gf A^m - 2 i \gf D 
\big)~.  
} 
Under the five dimensional super gauge transformations 
$\gd_\gL V = \gL + \bgL$ and $\gd_\gL S = \sqrt 2 \der_5 \gL$ 
the Chern--Simons superfields transform as 
\equ{
\arry{l}{\dsp 
\gd_\gL  \gO \ = - \frac 14 D^\ga (\gL W_\ga) 
- \frac 14 \bD_\dga (\bgL \bW^\dga)~, 
\\[2ex] \dsp 
\gd_\gL Y_\ga =  \frac i4 \bD^2 
\big( 
V_5 D_\ga \gL - D_\ga V_5 ( \gL + \bgL) 
+ 2 \der_5 \gL D_\ga V 
\big)~,
} 
\labl{gaugeSCS}
} 
where we have exploited the chirality of various superfields.

The Chern--Simons multiplets $\gO$ and $Y^\ga$ will play an important
role in determining the supersymmetric Green--Schwarz theory in
five dimensions. Before proceeding to this discussion in the next subsection
let us mention two interesting uses of these multiplets. 
First, equation \eqref{defCSsuper} implies that the super Yang--Mills
action can be written in terms of $\Omega$ as 
\equ{
S_{SY\! M} = \int \d^5 x \int \d^4 \gth (-2\gO + V_5^2) 
= 
 \int \d^5 x \bigg[
 \int \d^2 \gth \frac 14 W^\ga W_\ga + \text{h.c.}
 + 
\int \d^4 \gth \, V_5^2 
\bigg]~,
\labl{SYM}
}
with the component form of this action given in \eqref{AcSVcomp}. In
the following section we will encounter the combination 
$2 \gO -V_5^2$ frequently.

Secondly, both Chern--Simons 
multiplets can be used to give a compact representation of supersymmetric
Chern--Simons interactions in five dimensions. In particular, the mixed
Chern--Simons interaction involving two gauge fields $A_M$ and $\cA_M$ can 
be conveniently written as  
\equa{
S_{CS\, mix} &  = \int \d^5x  \bigg[ 
\int \d^2 \gth -\frac i2 Y^\ga \cW_\ga + \text{h.c.} + 
\int \d^4 \gth~2 \cV_5 \big( 2 \gO - V_5^2 \big) 
\bigg]  
\labl{mixSCS} \\[1ex] 
 & \supset \int\!  \d^5 x \Big[ 
- \frac 1{12} \ge^{MNPQR} \go_{MNP} \cF_{QR} 
+ \gvf \Big( 
\frac 12 F_{MN} F^{MN} 
+ \der_M \gf \der^M \gf 
- \big( D + \der_5 \gf \big)^2 
- 2\bF F 
\Big) 
\non \\[1ex]  
& \qquad \qquad 
+ 2 \gf \Big( 
\frac 12 F_{MN} \cF^{MN} 
+ \der_M \gf \der^M \gvf 
- \big( D + \der_5 \gf \big) \big( \cD + \der_5 \gvf \big)
- \bcF F -  \bF \cF 
\Big) 
\Big]~. 
\labl{mixSCScomp} 
} 
By comparing with the prepotential action \eqref{AcPreV} we infer that
this mixed Chern--Simons interaction can also be obtained from the
prepotential $\cP_{CS\, mix} = - \gvf \gf^2$.

Let us close this subsection with a short
discussion on the generalization to multiple vector multiplets. The 
Chern--Simons superfield $\gO(V,V')$ will be used to make contact
with the holomorphic prepotential language of $N\!=\!2$ supersymmetry in
four dimensions.  The five dimensional Chern--Simons term (the last
term of the first line of equation \eqref{AcPreV}) can be rewritten using  
\equ{
\int \d^4 \theta \, 
\big( \cV^a D^\ga \der_5 \cV^b - D^\ga \cV^a
\der_5 \cV^b \big) \cW^c_\ga + \text{h.c.} = 
8 \int \d^4 \theta \, \partial_5 \cV^b \, \gO(\cV^a, \cV^c)~. 
\label{ConvgO}
}
It is not difficult to see that this gives a five
dimensional realization of a generalized Chern--Simons term containing
two gauge fields and a field strength. In the four dimensional $N\!=\!2$
context, the existence of such terms was observed a long time ago 
\cite{wlp} (more recently see also \cite{aafl,wst}). Their presence in
$N\!=\!1$ theories was pointed out by using anomaly type arguments in
phenomenological models \cite{akt}, in
the deconstruction of $\U{1}$ supersymmetric gauge theories \cite{dfp}, 
flux compactifications and generalized Scherk--Schwarz reductions
\cite{afl}. The general superspace
description of the Lagrangian and couplings was also recently worked out 
in \cite{afl}.

Using \eqref{ConvgO} we can rewrite \eqref{AcPreV} in a form that more
closely resembles the four dimensional $N\!=\!2$ action for vector
multiplets  
\equ{
S_{\cV}  = 
 \int \!\!  \d^5 x \int \!\!\d^4 \gth   
\Big[ \cK (\cS, \bar{\cS}) 
+ \frac {2 \sqrt{2}}{3} {\cF}_{abc}\,  \partial_5 \cV^b \gO(\cV^a, \cV^c)
\Big] 
+  \int \!\!  \d^5 x 
\bigg[ 
\int \d^2 \gth \frac 14  
\cF_{ab}( \cS) \cW^{a\ga} \cW^b_\ga 
 \\[1ex]
+ \int \!\!\d^4 \gth   
\frac 12 {\cF}_{ab}(\cS )  
\Big( \partial_5\cV^a \partial_5 \cV^b 
+ \sqrt{2} \cV^a  \partial_5 (\cS^b + \bar{\cS}^b)
\Big)
+ \text{h.c.} 
\bigg]~, 
\nonumber
}
where the K{\"a}hler potential $\cK (\cS, \bar{\cS})$ is defined by the
usual expression in terms of the holomorphic four dimensional prepotential
${\cF}(\cS)$: 
\equ{
 \cK (\cS, \bar{\cS})   = 
\frac {1}{2} \big( \cS^a {\bar{\cF}}_{a}(\bcS)  + 
\bar{\cS}_a {\cF}^{a}(\cS) \big)~, 
\qquad 
\cF(\cS) = \frac 12 \cP(\sqrt 2 \cS)~. 
\label{v5}  
} 
Since we are describing a five dimensional supersymmetric gauge
theory the holomorphic prepotential $\cF(S)$ is determined by 
the real prepotential $\cP(\gvf)$ in five dimensions defined in
\eqref{RealPre}. Consequently,  the coefficients of the holomorphic
prepotential $\cF(S)$ are real rather than complex. 


\subsection{Supersymmetrizing the dual Green--Schwarz theory}
\labl{sc:SusyDualGS}

As in our discussion of the bosonic Green--Schwarz mechanism in
section \ref{sc:DualGS},  we first focus on a hybrid  formulation of the
supersymmetric Green--Schwarz anomaly cancellation, which can serve as
a basis for both the vector or tensor multiplet description. To
construct the supersymmetrization of the bosonic Green--Schwarz
action \eqref{GSdual} we decompose the supersymmetric Green--Schwarz
action as 
\equ{
S_{SGS} = S_{SYM} + S_{STV} + S_{Bianchi} + S_{bdy} + S_{\go^2}~. 
\labl{SGS}
}
The actions $S_{STV}$ and $S_{SYM}$, given in \eqref{AcDualSusy} and
\eqref{SYM}, respectively, correspond to the
supersymmetrizations of the first three terms of \eqref{GSdual}. The
other parts of this action, $S_{Bianchi}$, $S_{bdy}$ and $S_{\go^2}$,
contain the supersymmetric extensions of the terms 
$\d \cB_2 * \go_3(A)$, $-\d \cB_2 X_2(A)$ and 
$\frac 12 * \go_3(A) \go_3(A)$ in \eqref{GSdual}, respectively.

The full superspace expressions and their bosonic reductions can be
obtained. The supersymmetric extension of the $\d \cB_2 * \go_3$ term is
\equa{
S_{Bianchi} & =
\int \d^5 x \bigg[ \int \d^2 \gth 
\frac 12 \cT_5^\ga Y_\ga 
+ \text{h.c.}  +  
\int \d^4 \gth~2 \cL (2 \gO - V_5^2) 
\bigg] 
\labl{bianchi} 
\\[1ex]
& \supset 
\int \d^5 x  \Big[ 
- \frac 16 \cH_{MNP} \go^{MNP} 
- \frac 16 \ge^{MNPQR} \gf \cH_{MNP} F_{QR} 
+ \frac 12 \gs F_{MN} F^{MN} 
\non \\[1ex]  & 
\qquad \qquad 
+ \gs \der_M \gf \der^M \gf 
- \gs \big( D + \der_5 \gf \big)^2 
- 2 \gs \bF F 
+ \cD_T \big( A_5 D - \der_m \gf A^m \big) 
\Big]~.  
\labl{bianchicomp} 
} 
The term $\cL V_5^2$ is separately gauge invariant, but it is required
to obtain a five dimensional Lorentz invariant expression. The
supersymmetrization of the boundary term 
$- \d \cB_2 X_2(A)$ is straightforward because there we only have
$N\!=\!1$ supersymmetry in four dimensions: 
\equ{
S_{bdy} = \left. 
\gx_I \int \d^4 x 
\int \d^2 \gth \frac i4 \cT^\ga W_\ga(V) 
\right|_{I} + \text{h.c.}  
\supset   
\left. \frac 12 \gx_I  \int \d^4 x 
\Big[ \frac 14 \ge^{mnkl} \cB_{mn} F_{kl}  - \gs D \Big] 
\right|_{I}.
\labl{bound} 
}
Finally, the supersymmetrized version of the
$\frac 12 * \go_3 \go_3$ interaction is given by 
\equa{
S_{\go^2} & = \int \d^5 x \bigg[ 
\int \d^2 \gth \frac 14 Y^\ga Y_\ga 
+ \text{h.c.} 
- \int \d^4 \gth \big( V_5^2 - 2 \gO \big)^2   
\bigg] 
\labl{omega2}
\\[1ex]
& 
\supset 
 \int \d^5 x \Big[
- \frac 1{12} \go_{MNP} \go^{MNP} 
- \frac 16 \ge^{MNPQR} \gf \go_{MNP} F_{QR} 
+ \frac 32 \gf^2 F_{MN} F^{MN} 
\non \\[1ex] 
& \qquad  \qquad 
+ 3 \gf^2 \der_M \gf \der^M \gf
- 3 \gf^2 \big( D + \der_5 \gf \big)^2 
- 6 \gf^2 \bF F 
+ \frac 12 \big( A_5 D - \der_m \gf A^m \big)^2 
\Big]~.
\labl{omega2comp}
}

In a supersymmetric theory the gauge anomaly lifts to a full super
gauge anomaly i.e. the effective action with all charged boundary
and bulk matter multiplets integrated out transforms as
\cite{Grisaru:1985hc,Arkani-Hamed:1997mj} 
\equ{
\gd_\gL \gG_{eff} = 
\sum_I \gx_I \int \d^5 x \int \d^2 \gth~\gL  W^\ga W_\ga \gd(y - I) 
+ \text{h.c.}~. 
\labl{SanomgGeff} 
}
Since this variation does not involve any Green--Schwarz superfields
and it is quadratic in the vector multiplet $V$, we obtain the
following super gauge transformations 
\equ{
\gd_\gL \widehat \cW_\ga = i \gd_\gL \cT_{\!5\, \ga} = - i \gd_\gL Y_\ga~, 
\qquad 
\gd_\gL \widehat \cV_5 = \gd_\gL \cL = 2 \gd_\gL \gO~, 
\labl{AnomGaugecWcV5}
} 
by requiring that the action $S_{SGS}$ cancels the anomalous
variation. 
Moreover, using the super gauge transformation $\gd_\gL Y_\ga$ and 
$\gd_\gL \gO$ of the Chern--Simons superfields, given in
\eqref{gaugeSCS}, we infer that the multiplets $\cT_\ga$ and $\cU$
transform as    
\equ{
\gd_\gL  \cT_\ga = 
2i \gL W_\ga(V) = 
- \frac i2 \bD^2\big( \gL D_\ga V \big)~, 
\qquad 
\gd_\gL \cU = i ( \gL - \bgL) V_5~. 
\labl{TUgauge}
}
These equations constitute the supersymmetric version of the 
anomalous gauge transformations of the anti--symmetric tensor
$\cB_{MN}$ given in \eqref{GaugeTensors}. The first equation in
(\ref{TUgauge}) contains
the transformation of $\cB_{mn}$ while the second equation
contains the transformation of $\cB_{m5}$. With these gauge 
transformations it follows that the
supersymmetric Green--Schwarz action is super gauge covariant  
\equ{
\gd_\gL S_{SGS} = 
- \int \d^4 x \int \d^2 \gth~\gx_I \gL W^\ga W_\ga \gd(y-I) + \text{h.c.}~,
\labl{varSGS}
}
and this contribution precisely cancels the anomalous variation 
\eqref{SanomgGeff} of brane and bulk fermions in a supersymmetric theory.
We stress that this gauge transformation
is independent of the choice of using the supersymmetrized two-- or
one--form tensor formulation of the Green--Schwarz theory.

Having obtained all the parts which constitute the supersymmetrization
of the action \eqref{GSdual}, we can now proceed as in section
\ref{sc:SusyDual} to eliminate some of the superfields to obtain the
supersymmetric analogy of the Green--Schwarz mechanism formulated
using a five dimensional one-- or two--form. 

\subsection{Vector multiplet formulation
}
\labl{sc:SusyVecGS}

As in section \ref{sc:SusyDual} eliminating $\cU$ and $\cT_\ga$
gives rise to constraint equations which can be solved in terms of a
vector multiplet $\cV$ and a chiral multiplet $\cS$.  Because of the
additional Green--Schwarz interactions the solutions \eqref{solcW} and
\eqref{solcV5} are modified to 
\equ{
\widehat \cW_\ga = \cW_\ga (\cV) - i Y_\ga~, 
\qquad 
\widehat \cV_5 = \cV_5 - V_5^2 +2 \gO - \gx_I V \gd(y-I)~, 
\labl{solhatWV5}
} 
where $\cS$ is contained in the gauge invariant superfield 
$\cV_5$ introduced in
\eqref{superFieldStrength}. The super gauge transformations
\eqref{AnomGaugecWcV5} imply that 
\equ{
\gd_\gL \cV = 0~, 
\qquad 
\gd_\gL \cS = \sqrt 2 \sum_I \gx_I \gL \gd(y - I)~. 
}
These variations are compatible with \eqref{anomOne} where the four
dimensional vector multiplet is invariant, and only the chiral superfield
$\cS$ containing $\cA_5$ has a singular anomalous gauge transformation.

The supersymmetric one--form description of the Green--Schwarz
mechanism can be encoded with these expressions as 
\equa{
S_{SGS1} & = \int \d^5 x \bigg[ 
\int \d^2 \gth \frac 14 \Big( W^\ga W_\ga + 
\widehat \cW^\ga \widehat \cW_\ga + Y^\ga Y_\ga \Big) 
+ \text{h.c.} 
\non \\[1ex] 
&\qquad\qquad + \int \d^4 \gth \big( V_5^2 + \widehat \cV_5^2 
- (V_5^2 - 2 \gO)^2 \big) \bigg]~. 
}
The structure of this action is very similar to \eqref{Vaction}, since
it is also written as a sum of squares. 
In particular it also contains a singular $\gd(0)$ term. To see this
explicitly we substitute the relations \eqref{solhatWV5} and 
rewrite the action as 
\equa{
S_{SGS1} &= \int \d^5 x \bigg[
\int \d^2 \gth  \frac 14 \Big( W^\ga W_\ga + \cW^\ga \cW_\ga \Big) + \text{h.c.} + 
\int \d^4 \gth \Big( V_5^2 + \cV_5^2 \Big) 
\bigg] 
\non \\[1ex]
&+ \int \d^5 x \Big[
\int \d^2 \gth - \frac i2 Y^\ga \cW_\ga + \text{h.c.} + 
\int \d^4 \gth~2 \cV_5 \big( 2 \gO - V_5^2 \big) 
\Big]  
\non \\[1ex]
&+\left.
\frac 12\, \gx_I \int \d^4 x \int \d^4 \gth \Big[ 
-2 V\big( \cV_5 + 2 \gO - V_5^2 \big) + \gx_I \gd(0) V^2 
\Big] \right|_I~. 
\labl{SGS1}
}
The second line of this equation is identical to \eqref{mixSCS}
and consequently represents a supersymmetric mixed--Chern--Simons term.
The bulk actions of the vector multiplets, described by the
first two lines of \eqref{SGS1}, can be obtained from the 
prepotential 
\equ{
\cP_{SGS1} =  \frac 12 \gf^2 + \frac 12 {\gvf}^2 - \gvf \gf^2~, 
\label{PreSGS1} 
} 
using \eqref{AcPreV}. Hence, just like in the non--supersymmetric
version of the Green--Schwarz theory in the vector multiplet
formulation (see below \eqref{Vaction}), this supersymmetric action
only transforms in the required way, \eqref{varSGS}, if the multiplet
$\cV$ vanishes at the boundaries.

The last line of \eqref{SGS1} reads 
\equa{
&\left. 
\frac 12\, \gx_I \int \d^4 x \int \d^4 \gth \Big[ 
-2 V\big( \cV_5 + 2 \gO - V_5^2 \big) + \gx_I \gd(0) V^2 
\Big] \right|_I 
\non \\[1ex]
&\supset \gx_I  \int \d^5 x \Big( 
- D (\gvf - \gf^2) + A^m ( \cF_{m5} - 2 \gf F_{m5}) 
- \frac 12 \gx_I \gd(0) A_m A^m 
\Big) \gd(y - I)~.
}
The component form of the singular term involving the $\gd(0)$ is
exact; no fermionic terms were dropped.  Comparing to the bosonic
Green--Schwarz action in the vector formulation \eqref{Vaction}, we
see that this singular term and the $A^m \cF_{m5}$ term are correctly
reproduced. Therefore the analysis of the interpretation of the
$\gd(0)$ boundary mass seems identical to the bosonic discussion in 
subsection \ref{sc:U1mass}. Moreover, one can show that the mass
spectrum is supersymmetric. By eliminating the auxiliary $D$ field one
obtains mass terms for $\gvf$ that lead to the same recursive mass
eigenvector equation as \eqref{EVrec}. In addition, we have performed
an investigation of the boundary Dirac mass term resulting from the term
$-2 V \cV_5$ in the last line of \eqref{SGS1} in Wess--Zumino gauge,
and confirmed that the mass spectrum is identical to that of the
bosons derived in subsection \ref{sc:U1mass}. The calculation of
the fermion mass spectrum is quantitatively similar to that obtained
in a different context for boundary gravitino masses~\cite{bfz}.
Since in the supersymmetric context we are restricting ourselves to the
bosonic reduction, we do not report this calculation explicitly here.

Notice that the superspace term $V \gO$ does not give a
contribution as it is a total derivative in superspace
\cite{Cecotti:1987nw} 
\equ{
V \gO(V, V') = - \frac 18 D^\ga( V^2 W_\ga(V') ) + \text{h.c.}~,
\labl{totSder}
}
for any two vector multiplets $V$ and $V'$. This is the supersymmetric
version of the statement below \eqref{Vaction} that $\go_3(A)X_2(A)$ drops
out of the action. 

However, if we have
multiple vector multiplets, we encounter combinations like $V' \gO(V,
V')$ as well. These are not total 
derivatives in superspace and give rise to generalized Chern--Simons
interactions as discussed in Appendix~\ref{sc:MultiU1}.

According to \eqref{AcPreV} the kinetic terms of the vector multiplets
are determined by the field metric
\equ{
\cP_{ab} = 
\pmtrx{ 1 - 2 \gvf & -2 \gf \\[1ex] -2 \gf & 1 }~, 
\qquad 
\det \big( \cP_{ab} \big) = 1 -2 \gvf - 4 \gf^2~,  
\labl{Metric}
} 
derived from the prepotential \eqref{PreSGS1} (dropping the
subscript). The eigenvalues of the field metric are
\(
\gl_\pm = 1 - \gvf \pm ( \gvf ^2 + 4 \gf^2)^{1/2}~. 
\)
Notice that the eigenvalue $\gl_+$ is positive; in fact $\gl_+ \geq 1$. 
The other eigenvalue $\gl_-$ changes sign where the Hessian 
$\det \big( \cP_{ab}\big)$ vanishes. Since some kinetic terms vanish
there, the theory develops a Landau pole and near this region the
anomalous $\U{1}_A$ gauge theory is strongly coupled. This signals
that the effective field theory description probably breaks down.

\begin{figure}
\begin{center}
\raisebox{0ex}{\scalebox{.6}{\mbox{\input{Landau.pstex_t}}}}
\end{center}
\caption{
This plot indicates the regions where the Hessian 
$\det \big(\cP_{ab}\big)$ and the eigenvalue $\gl_-$ are positive and
negative. At the parabola $\gvf(\gf) = \frac 12 - 2 \gf^2$ they vanish
and a Landau pole arises. 
\labl{fg:LandauPole}}
\end{figure}


\subsection{Tensor multiplet formulation
}
\labl{sc:SusyTenGS}

The tensor formulation is obtained by eliminating the superfields 
$\smash{\widehat \cW_\ga}$ and $\smash{\widehat \cV_5}$ with their
algebraic equations of motion. In fact the expressions for these
superfields are identical to the ones given in \eqref{solcWcV5} of
section \ref{sc:SusyDual}. Therefore the tensor multiplet form of the
Green--Schwarz action is given by 
\equa{
S_{SGS2}  &  = 
\int \d^5 x \bigg[
\int \d^2 \gth \Big( \frac 14 W^\ga W_\ga + 
\frac 14 \big( \cT_5^\ga + Y^\ga \big)  \big( \cT_{\!5\,\ga} + Y_\ga \big) 
+  \frac i2  \gx_I \cT^\ga W_\ga \gd(y - I) \Big) 
+ \text{h.c.} 
\non \\[1ex]  & \qquad \qquad 
+ \int \d^4 \gth \Big( V_5^2 - \big( \cL -2 \gO + V_5^2 \big)^2 \Big) 
\bigg]
\labl{SGS2}  \\[1ex] & 
 \supset 
 \int \d^5 x \Big[
- \frac 1{12} (\cH_{MNP}+\go_{MNP})(\cH^{MNP}+\go^{MNP}) 
- \frac 16 \ge^{MNPQR} \gf (\cH_{MNP}+ \go_{MNP}) F_{QR} 
\non \\[1ex] & \qquad \qquad 
 - \frac 12 \der_M \gs \der^M \gs
- \Big( \frac 12 - \gs - 3 \gf^2\Big) 
\Big(
\frac 12 F_{MN} F^{MN} 
+ \der_M \gf \der^M \gf 
- \big( D + \der_5 \gf \big)^2 - 2 \bF F 
\Big) 
\non \\[1ex] & \qquad \qquad 
+ \frac 12 \big( \cD_T + A_5 D - \der_m \gf A^m \big)^2 
+ \gx_I \Big( \frac{1}{4} \ge^{mnkl} \cB_{mn} F_{kl} - \gs D \Big) 
\gd(y-I)
\Big]~.  
\labl{SGS2comp} 
} 
Notice that a delta function only appears in the middle term of (\ref{SGS2}).
As in the non--supersymmetric case we see that no $\gd(0)$ terms
appear on the boundaries in the tensor multiplet formulation of the
Green--Schwarz theory.

We have also given the complete Lorentz invariant bosonic part of the 
Lagrangian by combining our earlier results (\ref{AcSTcomp}), 
(\ref{bianchicomp}), (\ref{bound}) and (\ref{omega2comp}). Notice that
the equation of motion of the auxiliary field $\cD_T$ is modified, 
but since it still appears as a complete 
square it can eliminated without leaving any trace. This is not
the case anymore for the auxiliary field $D$ because of the presence
of the last term in the bosonic reduction \eqref{SGS2comp}.  We will 
discuss the consequences of this in section \ref{sc:Pheno}.


\subsection{Supersymmetric duality relations}
\labl{sc:susyDual}

By combining \eqref{solcWcV5} and \eqref{solhatWV5} 
for $\smash{\widehat \cW_\ga}$ and $\smash{\widehat \cV_5}$ the supersymmetric generalizations 
of the duality relations \eqref{onetwodual} are found to be
\equ{
\arry{c}{  \dsp 
i \cT_{5\, \ga} = 
i\big( \der_5 \cT_\ga + \cW_\ga(\cU) \big) = \widehat \cW_\ga = 
\cW_\ga (\cV) - i Y_\ga, 
\\[1ex] \dsp 
\cL = 
 \frac i4  (D^\ga\cT_\ga - \bD_\dga \bcT^\dga) = 
\widehat \cV_5 = \cV_5 - V_5^2 +2 \gO - \gx_I V \gd(y-I)~. 
}\labl{dualrel}
}
By taking the appropriate components of these superfield expressions,
one retrieves the one/two--form duality. In particular, the bosonic
component of the first equation in \eqref{dualrel} gives rise to three relations 
\equ{
\arry{c}{\dsp 
\cH_{mn5} + \go_{mn5}  = 
\frac 12 \ge_{mnkl} \big( \cF^{kl} - 2 \gf F^{kl} \big)~, 
\\[2ex]  \dsp 
\cD = - \der_5 \gs + 2 \gf D~, 
\qquad 
\cD_T = - A_5 D +  \der_m \gf A^m~, 
}
\labl{Hmn5dual}
}
where we have dropped all fermionic contributions for simplicity. 
The latter two relations are the equation of motions of the auxiliary
fields $\cD$ and $\cD_T$ in the vector and tensor formulation of the
supersymmetric Green--Schwarz mechanism.

The first equation in \eqref{Hmn5dual} is a part of the (4+1)
dimensional decomposition of the duality between the five dimensional
tensor $\cB_{MN}$ and vector $\cA_M$. The other part of this 
decomposition
\equ{
\frac 16 \ge_{pmnk} \big( \cH^{mnk} + \go^{mnk} \big) = 
\cF_{p5} - 2 \gf F_{p5} - \gx_I A_p \gd(y - I)~,
\labl{Hmnkdual} 
}
is obtained by taking the vector ($\frac 12 [D_\gb, \bD_\dgb]$)
component of the second equation in \eqref{dualrel}. These two duality
relations in (4+1) dimensional notation can be combined in form
notation to give 
\equ{
- * \d \cB_2 =  \d \cA_1 + * \go_3 - X_2(A) - 2 \gf \d A_1~. 
\labl{onetwodualsusy}
}
Notice that compared to the duality relation \eqref{onetwodual} of the
non--supersymmetric Green--Schwarz mechanism discussed in section
\ref{sc:DualGS}, the supersymmetrization has introduced a new term 
$ -2 \gf \d A_1$.  In addition to this bosonic term there are various
fermionic modifications to this duality relation that we have dropped
in our bosonic reduction. This shows that even in the bosonic reduction
supersymmetry leads to a modification of the duality between the
vector and tensor description of the Green--Schwarz action.

It is instructive to determine the content of the other bosonic
components of the second expression in (\ref{dualrel}). The lowest
and the $D^2$--components give rise to 
\equ{
\gs = \gvf - \gf^2 = \frac{\der \cP_{SGS1}}{\der \gvf}~, 
\qquad 
\cF = 2 \gf F~.
\labl{ScalarDual}
}
The expression for $\gs$ reproduces the duality relation between
scalar components of the tensor and vector multiplet formulation in five
dimensions~\cite{Seiberg:1996bd}. It is an exact relation: No
fermionic terms have been dropped. The second relation is the
algebraic equation of the auxiliary field $\cF$, as can be seen by combining
the actions \eqref{AcSVcomp} and \eqref{mixSCScomp}. Finally the $D$--term
component of the second relation in \eqref{dualrel} can be cast in the
form 
\equ{
\der_M \der^M \gs + \frac 12 F_{MN} F^{MN} + \der_M \gf \der^M \gf
- \big(D + \der_5 \gf \big)^2 - 2 \bF F - \gx_I D \gd(y - I) = 0~, 
\labl{EMgs}
 }
using the auxiliary field equation \eqref{Hmn5dual} for $\cD$.
As can be seen from \eqref{SGS2} this is the equation of motion of the scalar
field $\gs$. The fact that the supersymmetric duality relations
encode various equations of motion is not surprising in the light of
the observation below simplest realization of the vector/tensor
duality \eqref{VTduality}, that this duality contains the equations of
motion for $\cB_2$ and $\cA_1$.

At the end of subsection \ref{sc:SusyVecGS} we commented on the
conditions for the appearance of Landau poles in the vector
formulation. As a Landau pole constitutes a physical effect, it should
be independent on which formulation one chooses to work in. Indeed, also
in the tensor multiplet formulation we encounter the possibility of
Landau poles. If in \eqref{SGS2comp} the field combination 
$1 - 2 \gs - 6 \gf^2$ vanishes, the kinetic terms of the vector
multiplet disappear. Upon using the scalar duality relation (first
equation in \eqref{ScalarDual}) we see that this happens precisely
when the Hessian $\det \big( \cP_{ab}\big)$, given in \eqref{Metric},
is zero. Observe that in both the vector and the tensor formulation
the anomalous $\U{1}_A$ vector multiplet has kinetic mixing with the
Green--Schwarz fields.

\section{Phenomenological aspects}
\labl{sc:Pheno}

In this section we investigate some simple consequences of
supersymmetric five dimensional theories with an anomalous matter
spectrum. As before we consider the theory on an interval or 
equivalently the orbifold $S^1/\Intr_2$. The matter may be either bulk
hypermultiplets or chiral multiplets on the boundaries. All matter is
only charged under the anomalous $\U{1}_A$, but does not couple
directly to the Green--Schwarz superfield, irrespectively of whether it
is described by the vector or tensor multiplet formulation. The
$\Intr_2$ parities of the various multiplets that determine the
orbifold boundary conditions are given in Table~\ref{tb:Parities}. 
To fix our notation we first briefly recall how charged bulk hyper and 
brane chiral multiplets are described using $N\!=\!1$ superfields.
This will enable us to investigate vacuum solutions of anomalous
theories using both the vector and tensor multiplet formulation of the
Green--Schwarz theory developed in section \ref{sc:SusyGS}.  
In particular we will focus on the role of the duality
relations found in subsection \ref{sc:susyDual}.

\begin{table}
\[
\arry{c | c | c | c }{
\text{even} & 
\gF_+ & V & \cS 
\\[1ex] \hline & & & \\[-2ex] 
\text{odd} & 
\gF_- & S & \cV 
}
\]
\caption{ 
The even and odd superfields under the $\Intr_2$ orbifold action. 
\labl{tb:Parities}}
\end{table}

Bulk hypermultiplets with charge matrix $q_b$ can be described by the
chiral $N\!=\!1$ multiplets $\Phi_+$ and $\Phi_-$, where the subscript refers
both to the $\Intr_2$ parity of the multiplets and the sign of their
charges.  The charged hypermultiplet action reads 
\equa{
S_{hyper}  & = \int \d^5 x \bigg[ 
\int \d^2 \gth \gF_- \big( \der_5 + \sqrt 2 q_b S) \gF_+ + \text{h.c.} +
\int \d^4 \gth \ \bgF_\pm e^{\pm 2q_b V} \gF_\pm
\bigg]  
\labl{hyp} 
\\[1ex] 
& \supset 
\int \d^5 x \Big[ 
- \big|  \der_M \gvf_\pm \pm i q_b A_M \gvf_\pm \big|^2 
- \gf^2 \bgvf_\pm q_b^2 \gvf_\pm 
\pm \bgvf_\pm q_b \gvf_\pm ( D + \der_5 \gf ) 
\non \\[1ex] 
& \qquad \qquad 
+ \sqrt 2 ( \gvf_- q_b \gvf_+ F + \text{h.c.} ) 
+ 
\big| \bF_\mp \pm \big(\der_5 \pm q_b (\gf + i A_5) \big) \gvf_\pm \big|^2 
\Big]~,  
\labl{hypcomp}
} 
where summation over both chiral multiplets is implied. For the chiral
matter $\gF_I$ on the boundaries one has instead  
\equa{
S_{chiral} &= \int \d^5 x \int \d^4 \gth \Big[ 
 \bgF_I e^{ 2 q_I V} \gF_I
\Big] \gd(y - I)
\labl{chiral} 
 \\[1ex] &  
\supset 
\int \d^5 x \Big[ 
- \big|  \der_m \gvf_I + i q_I A_m \gvf_I \big|^2 
+ \bgvf_I q_I \gvf_I  D 
\Big] \gd(y - I)~.
\labl{chiralcomp}
} 
Finally brane localized Fayet--Iliopoulos terms are given by 
\equ{
S_{FI} = 
 \int \d^5 x \int \d^4 \gth \ \gx_I \kappa V \ \gd(y - I) 
= \int \d^5 x \ \gx_I \kappa D \ \gd(y - I)~,
}
where we have introduced a constant $\kappa$ that is proportional to the
cut--off scale if one assumes that these Fayet--Iliopoulos terms are generated at
one--loop, and the coefficients $\gx_I$ are given by
\eqref{FIcoef}. (We will ignore subtle issues that the $D$--terms
induced by bulk loops extend into the bulk
\cite{Scrucca:2001eb,GrootNibbelink:2002wv,GrootNibbelink:2002qp}.)

\subsection{Vacuum structure of the Green--Schwarz theory}
\labl{sc:Vacuum}

In the vector multiplet formulation of the Green--Schwarz theory we
find the following equations for the auxiliary fields
\equ{
(1 - 2 \gvf) (D + \der_5 \gf) - 2 \gf (\cD + \der_5 \gvf) 
\pm  \bgvf_\pm q_b \gvf_\pm 
+ \big( \gx_I ( \kappa - \gvf + \gf^2 ) + \bgvf_I q_I \gvf_I \big)
\gd(y-I) =  0, 
\non \\[1ex] 
\cD + \der_5 \gvf - 2 \gf(D + \der_5 \gf) = 0~, 
\quad 
\cF - 2 \gf F = 0, 
\labl{BPSvector}
\quad 
(1 - 2 \gf )\bF - 2 \gf \bcF + \sqrt 2 \gvf_- q_b \gvf_+ = 0~. 
} 
In order for the background to preserve supersymmetry the auxiliary
fields should have zero VEVs. In particular, from the first equation 
on the second line we see that this implies that 
$\langle\gvf \rangle - \langle \gf \rangle^2$ is constant.

The tensor formulation has only two auxiliary fields ($D$ and $F$) 
relevant for the determination of the background, since the remaining
auxiliary field $\cD_T$ is expressed in terms of gauge fields 
(see \eqref{Hmn5dual}) that do not acquire VEVs. The $D$ and $F$ 
auxiliary field equations 
\equ{
(1 - 2 \gs - 6 \gf^2) (D + \der_5 \gf) 
\pm  \bgvf_\pm q_b \gvf_\pm 
+ \big( \gx_I ( \kappa - \gs ) + \bgvf_I q_I \gvf_I \big)
\gd(y-I) = 0~, 
\non \\[1ex]
(1 - 2 \gs - 6 \gf^2) \bF + \sqrt 2 \gvf_- q_b \gvf_+ = 0~, 
\labl{BPStensor} 
} 
are equivalent to combinations of the relations in
\eqref{BPSvector}. Indeed, by eliminating $\cD$ and $\cF$ from those
equations and using the identification $\gs = \gvf - \gf^2$ one
obtains \eqref{BPStensor}. However, contrary to the auxiliary field  equations in
the vector formulation, the auxiliary field equations in the tensor formulation do
not imply that $\langle \gs \rangle$ is constant. In the tensor
formulation this conclusion is instead reached by considering the
equation of motion  \eqref{EMgs} for a supersymmetric background.

Let us consider two simple of examples of matter configurations to
illustrate some consequences of the supersymmetrized
Green--Schwarz theory in five dimensions. In the first example we take
a single chiral multiplet $\gF_0$ with charge $+1$ at the boundary
$y=0$. The local cubic sum of charges thus equals $\gx_0 = 1/(24\pi^2)$ and 
$\gx_\gp =0$. Notice that the mixed U(1)--gravitational anomaly is
automatically cancelled since this charge assignment trivially
satisfies the universality relation. Assuming that 
\(
\langle \gs \rangle = \langle \gf \rangle 
- \langle \gvf \rangle^2 
\)
is constant, we find that the first equation of \eqref{BPStensor} (or
the equivalent equations in \eqref{BPSvector}) leads to the conditions 
\equ{
\xi_0(\gk - \langle \gs \rangle) + \big| \langle \gvf_0 \rangle \big|^2 = 0~, 
\qquad 
\big(1 - 2 \langle \gs \rangle - 2 \langle \gf \rangle^2 \big) 
\langle \gf \rangle = 0~. 
}
We see that any value of the Fayet--Iliopoulos parameter $\gk$ and the
brane field VEV $\langle \gvf_0 \rangle$ can be compensated by an
appropriate VEV of $\langle \gs \rangle$.  The second equation implies
that $\langle \gf \rangle = 0$ is always the only solution. For 
\(
1 - 2 \langle \gs \rangle \geq 0 
\) 
two additional solutions arise with 
\(
\langle \gf \rangle =  \sqrt{(1-2 \langle \gs \rangle)/2} \ge(y). 
\)
This is because $\gf$ is odd and 
the solution will have to be proportional to the step function $\ge(y)$, 
defined by  $\partial_5\epsilon(y) =2(\delta(y)-\delta(y-\pi))$.

In the second example we consider a model with one bulk hypermultiplet
$(\Phi_+ , \Phi_-)$ of charge~$+3$, and single brane chiral multiplets
on both boundaries with equal charge $-1$. For the cubic sum of local 
charges we find $\gx \equiv \gx_0 = \gx_\gp = 25/(2\cdot24\pi^2)$. Again
these charges satisfy the universality relation and the mixed 
U(1)-gravitational anomaly is cancelled. In this case the auxiliary
field equations, \eqref{BPSvector} or \eqref{BPStensor}, can be solved by 
\equ{
\arry{c}{\dsp 
\big| \langle \varphi_+ \rangle \big|^2  = 
\big| \langle \varphi_- \rangle \big|^2 =0~, 
\quad  
\big| \langle \varphi_0 \rangle \big|^2
+ 
\big| \langle \varphi_\gp \rangle \big|^2
= 
2 \gx \big(\kappa - \langle \gs \rangle \big)~, 
\\[1ex] \dsp 
\big(1 - 2 \langle \gs \rangle - 2 \langle \gf \rangle^2 \big) 
\langle \gf \rangle = 
\frac 14 \Big( 
\big| \langle \varphi_0 \rangle \big|^2
-
\big| \langle \varphi_\gp \rangle \big|^2
\Big) \ge(y)~. 
}
\label{p3} 
}
Notice that the VEV $\langle \gs \rangle$ is determined by sum of the
brane VEVs $| \langle \varphi_0 \rangle |^2$ and 
$| \langle \varphi_\pi \rangle |^2$, while the amplitude of
the jump of $\langle \gf \rangle$ is set by their difference. 

In Ref.\ \cite{Abe:2002ps} bulk profiles for the field (that we call $\gf$)
have been investigated when the overall sum of Fayet--Iliopoulos terms
is non--zero, i.e.\ $\gx_0 + \gx_\gp \neq 0$, but under the
assumption that a similar analysis can be performed as for the case in
which the sum of charges vanishes globally. In particular, as emphasized by
these authors, the field(s) responsible for cancelling the 
global anomaly were not specified. Our starting point was precisely to
see how the local and global anomalies can be cancelled via the five
dimensional supersymmetric version of the Green--Schwarz theory. As
the above results show for the structure of supersymmetric vacua, an
additional scalar $\gvf$ (or $\gs$) is introduced, and the auxiliary
field  equation for $\gf$ is cubic rather than linear in
$\gf$. However, qualitatively the analysis in Ref.\ \cite{Abe:2002ps} 
seems to be confirmed by our investigation since this cubic
equation also leads to non--trivial profiles for $\langle\gf\rangle$.

\subsection{Possible MSSM applications}
\labl{sc:applc} 

The vacuum expectation values for the scalar field $\phi$ 
clearly depend on the particular model under consideration, and can 
play a crucial role in generating various hierarchies for 
supersymmetry breaking and fermion masses in the MSSM. To briefly 
illustrate these possibilities consider a model with one bulk hypermultiplet
$(\Phi_+ , \Phi_-)$ of charge~$+2$, and one brane chiral multiplet 
$\Phi_0$ at $y=0$, of charge $-1$ under the anomalous gauge symmetry.
As shown in the previous subsection the scalar field $\varphi_0$ will 
obtain a VEV, and $\phi$ will satisfy a cubic equation leading to a
solution with a nontrivial profile. 
Next suppose that in the MSSM all fields live
on the $y=0$ boundary, in addition to the brane chiral field $\Phi_0$. 
The Yukawa couplings must respect the (anomalous) gauge $\U{1}_A$ symmetry 
and are generically given by higher dimensional operators \cite{fn}. 
Besides the field $\Phi_0$ with $\U{1}_A$ charge~$-1$, we denote the 
$\U{1}_A$ charges of the superfields $(Q_i,U_j,H_2)$ by 
small letters $(q_i,u_j,h_2)$. These charges 
must be chosen such that the universality relation is satisfied and
the mixed U(1) non--Abelian anomalies are cancelled. 
The Yukawa interactions are then described by the superpotential
\be
W  = \lambda_{ij} \Big(\frac {\Phi_0}{M}\Big)^{q_i + u_j + h_2} 
Q_i U_j H_2~, 
\label{p4}
\ee 
where $M$ denotes the fundamental scale, 
$Q_i$ denotes the superfield containing the left--handed $SU(2)_L$ 
doublet quarks of the $i$th generation,  $U_j$ contains the right--handed
$SU(2)_L$ singlet up--quarks of the $j$th generation, and $H_2$ contains the
Higgs doublet responsible for up--type masses in the MSSM. Thus, 
when $\varphi_0$ obtains a VEV, the superpotential term (\ref{p4})
generates Yukawa couplings for the fermions, and $\varphi_0$ plays the role
of the Froggatt--Nielsen scalar field.

Supersymmetry must also be spontaneously broken and in four dimensions 
this occurs by adding the gauge invariant superpotential 
term $\lambda\Phi_0 \Phi_0 \Phi_+ $. However, the supersymmetry 
breaking scale is determined by the scale of the FI term, which 
is typically (at least in heterotic constructions) much larger than the 
TeV scale. Instead,
from a five dimensional perspective, supersymmetry can be broken by 
adding the gauge invariant superpotential term
$\lambda \Phi_0 \Phi_0 \Phi_+(y=0)$ on the $y=0$ boundary. 
It is indeed straightforward (but more tedious than the four dimensional
case) to check that the five dimensional F and D flatness conditions 
$F_0 = F_+ = F_- = D = \cD=0$ have no solution in this model. 
However, compared to four dimensions, the supersymmetry breaking scale 
will be suppressed by the volume of the compact dimension via the wave 
function of the bulk field $\Phi_+$, using the mechanism proposed in a 
different context in \cite{ddg,addm}. Thus, unlike in four dimensions,
the FI term in this five dimensional model can induce supersymmetry 
breaking at the TeV scale.
  
If instead the charged matter fields live on different boundaries
then fermion mass hierarchies can be generated in a different way.
For example, if $\Phi_0$ ($\Phi_\pi$) are charged matter boundary fields 
at $y=0$ ($y=\pi$) with $\U{1}_A$ charges $-1$ ($+2$), 
then the gauge invariant superpotential term that breaks supersymmetry will
involve the Wilson line operator  
\be
W_{\rm susy-breaking} = \lambda \ \Phi_0 \Phi_0 
e^{2\sqrt{2} \int_0^{\pi R} dy \ S (y)} \Phi_\pi~.
\label{s2}
\ee
The Wilson line operator can induce a large
hierarchy between the fundamental scale and the supersymmetry breaking
scale \cite{dfp}. In this case the classical field equations determining 
${\rm Re} S = (1/\sqrt{2}) \gf$  have a solution similar to (\ref{p3}). 
All quark and charged lepton fields should live on the $y=0$ boundary 
in order to avoid a (too large) suppression in the fermion masses arising 
from the Wilson line operator.
On the other hand, if neutrino singlets $N_i$ live on the 
$y = \pi$ boundary, then the extra suppression from the Wilson line
operator allows a nice way of getting very small Dirac neutrino masses.
The relevant neutrino mass operator can be written as
\be
\lambda ^{\nu}_{ij} \Big(\frac {\Phi_0}{M}\Big)^{l_i + n_j + h_2} 
e^{\sqrt{2} n_j \int_0^{\pi R} dy\, S (y)} L_i N_j H_2~.
\label{s3}
\ee
Together with (\ref{s2}) the superpotential term (\ref{s3}) predicts a 
relation between the neutrino mass scale and the supersymmetry breaking scale.
A similar mechanism in four dimensions was proposed in \cite{ahmsw,borzumati} 
by using a Kahler potential operator and the Giudice--Masiero mechanism. 
In contrast, in our case this relation is generated by the superpotential 
via a large suppression occurring from the geometric separation between the 
various charged fields that produces the Wilson line operator.

\section{Conclusions}
\labl{sc:concl}

In five dimensions the Green--Schwarz mechanism can be used to cancel
$\U{1}_A$ gauge anomalies. This mechanism relies on introducing a bulk
field whose gauge variation is responsible for anomaly cancellation on
the boundaries.  In five dimensions this field can either be described
by a rank two tensor field $\cB_2$ or a vector field $\cA_1$ that 
are dual to each other. This duality has been extended to the Green--Schwarz 
action where the anomalous $\U{1}_A$ gauge field $A_1$ interacts with 
either the (non--anomalous) vector field $\cA_1$ or the tensor field
$\cB_2$. In the vector formulation the Green--Schwarz action contains
singular boundary mass terms for the anomalous gauge field $A_1$, which are
absent in the tensor formulation.  However, we showed
that both formulations do give the same mass spectrum for
the anomalous $\U{1}_A$ gauge field. A general
characteristic of this spectrum is that the lowest
lying mode is massive when the global sum of charges is different
from zero. In the vector formulation this involved a cancellation of the
$\delta(0)$ boundary mass terms by a mixing between the anomalous and
Green--Schwarz gauge fields. In the tensor formulation there are no
(singular) boundary mass terms, and the Kaluza--Klein masses 
arise from  a boundary mixing term between the Green--Schwarz
tensor field and the anomalous gauge field. In this sense the tensor
formulation seems better defined because it is free of any singular terms at
the Lagrangian level.

A major part of our work has been to present the complete
supersymmetric extension of the five dimensional Green--Schwarz mechanism
using the four dimensional $N\!=\!1$ superspace. Each term in the
Green--Schwarz action has been supersymmetrized and written in a
manifestly $N\!=\!1$ supersymmetric form.  The components of the 
five dimensional gauge field $\cA_1$ become part of a vector multiplet 
$\cV$ and a chiral multiplet $\cS$. From these superfields, 
two super gauge invariant superfields $\cW_\ga(\cV)$ and $\cV_5$ are 
obtained.  In the dual formulation the supersymmetrization of the 
five dimensional tensor field $\cB_2$ is described by a 
chiral multiplet $\cT_\ga$ and a vector multiplet $\cU$. Also in this
dual formulation we have constructed two super gauge invariant 
combinations $\cT_{5\ga}$ and $\cL$ since, contrary to 
the four dimensional case, the supersymmetric interactions of $B_{mn}$
cannot be encoded by the linear multiplet $\cL$ only.
In addition the supersymmetric Chern--Simons three--form interactions
required the introduction of a new chiral Chern--Simons superfield $Y_\ga$
besides the Chern--Simons superfield $\Omega$ that would only be 
required in a four dimensional theory.

We have also supersymmetrised the vector/tensor duality and obtained the 
duality relations between the supermultiplets. The usual vector/tensor 
duality is now modified to include scalar degrees of freedom, aside from the
additional modifications due to fermionic terms. Furthermore, an 
interesting duality relation between the scalar components of the two
formulations is obtained. If these scalar components receive VEVs then
they can lead to Landau poles for the gauge field coupling.  In both
formulations we showed that the Landau poles occur on the same curve in
moduli space. In addition we have seen that the duality also plays 
a role in the equations that determine the supersymmetric vacua of the theory. 
In the vector multiplet formulation we encounter the same singular
boundary mass terms as in the non--supersymmetric case. Hence the Kaluza--Klein
mass spectrum for the anomalous $\U{1}_A$ gauge field in the 
vector and tensor formulation is identical to the one obtained in
the non--supersymmetric theory.  Moreover, both formulations lead to a finite
boundary Dirac mass terms for the gauginos. Thus, four dimensional
supersymmetry is preserved because the resulting fermionic
Kaluza--Klein mass spectrum is identical to that of the bosons.

Our analysis can be easily generalized to multiple anomalous $\U{1}_A$ gauge
groups. With this generalization more general Chern--Simons terms are
allowed that are not present in the single $\U{1}_A$ case. Anomalous $\U{1}_A$
symmetries have been extensively used for phenomenological purposes in
four dimensional models, such as for the fermion mass hierarchy and
supersymmetry breaking. In five dimensions these models can be further
enhanced by allowing supersymmetry breaking to occur at the TeV scale, and
to obtain extremely suppressed Yukawa couplings for Dirac neutrino masses.
These phenomenological aspects remain to be investigated further in
detail.
We expect the formalism developed in this paper to be useful as a starting
point for further theoretical studies of the Green--Schwarz mechanism in
higher dimensions. For example, generalizations to mixed $\U{1}_A$ non--Abelian
and mixed gravitational anomalies would be an interesting extension of
our analysis. We would also expect that six or more dimensions might
possibly be a more natural setting for the supersymmetric duality
relations derived in this paper. Another avenue would be to 
investigate the consequences of the Green--Schwarz interaction that we
found in this paper for one loop computations, in particular the
Fayet--Iliopoulos tadpoles. These are some of the interesting questions
that remain to be studied.

\section*{Acknowledgements}
We would like to thank S.\ Ferrara, C.\ Kounnas and M.\ Shifman 
for enlightening discussions, and M.\ Walter for useful comments on
the manuscript. 
The work of ED was supported in part by the RTN European Program
HPRN--CT--2000--00148. The work of TG and SGN is supported in part 
by the Department of Energy under contract DE--FG02--94ER40823 at 
the University of Minnesota. TG is also supported in part by a grant 
from the Office of the Dean of the Graduate School of the University of 
Minnesota. ED is grateful for the warm hospitality at the FTPI of the
University of Minnesota 
and the Theory Division of CERN during the early stages of this work.


\appendix 
\def\theequation{\thesection.\arabic{equation}} 

\setcounter{equation}{0}
\section{Deriving the defining equations for the Chern--Simons
superfields}
\labl{sc:SCSdefs}

To obtain the superfield definitions of the Chern--Simons multiplets
we will use a supersymmetric version of a Lagrange multiplier method. 
First, using a Lagrange multiplier one--form $\widetilde A_1$ the defining 
relation of the Chern--Simons three--form, $\d \go_3 = F_2^2$, can be 
obtained from the action 
\equ{
S = \int \widetilde A_1 \big( \d \go_3 - F_2^2 \big)
= 
\int \big(\widetilde F_2 \go_3 - \widetilde A_1 F_2^2 \big)~.
}
The (4+1) dimensional decomposition of this action, given by 
\equ{
S = \int \d^5 x~\ge^{mnpq} \Big( 
\frac 14 \widetilde F_{mn} \go_{pq5} - \frac 16 \widetilde F_{m5} \go_{npq} 
- \frac 14 \widetilde A_5 F_{mn} F_{pq} - \widetilde A_m F_{np} F_{q5} 
\Big)~, 
\labl{LagMultAction}
}
can then be supersymmetrized in a straightforward manner. Employing various
definitions given in the main text to identify the superfield terms, the
supersymmetrized version of (\ref{LagMultAction}) becomes
\equ{
S = \int \d^5 x \bigg[
\int \d^2 \gth \Big( 
\frac i2 \widetilde W^\ga Y_\ga - \frac 1{\sqrt 2} \widetilde S W^\ga W_\ga   
\Big) 
+ \text{h.c.} 
- \int \d^4 \gth \big( 
4 \widetilde V_5 \gO + 8 \widetilde V \gO(V_5, V) 
\big) 
\bigg]~. 
}
After a partial integration in superspace, and using 
$\int \d^2 \bgth = - \bD^2/4$, under a spacetime integral the action
can be rewritten as  
\equ{
\arry{l}{ \dsp 
S = \int \d^5 x \bigg[
\int \d^2 \gth 
\frac 1{\sqrt 2} \widetilde S \big(  \bD^2 \gO - W^\ga W_\ga \big) 
+ \text{h.c.}
\\[1ex]  \dsp 
\qquad\qquad 
+ \int \d^4 \gth~ \widetilde V 
\Big( 
\frac i2 \bD_\dga \bY^\dga - \frac i2 D^\ga Y_\ga 
- 4 \der _5 \gO   - 8 \gO(V_5, V) 
\Big) 
\bigg]~.
} 
}  
From this equation the defining properties, \eqref{defCSsuper} and
\eqref{defCSsuper5D}, of the Chern--Simons multiplets, $\gO$ and
$Y_\ga$, respectively, can be read off directly.

Furthermore it is straightforward to show that $\gO = \gO(V,V)$ is a solution 
of \eqref{defCSsuper}. On the other hand 
to prove that $Y_\ga$, given in \eqref{solY}, solves
\eqref{defCSsuper5D} it is useful to employ the following identities. For
two vector superfields $V$ and $V'$ we have 
\equ{
\frac 12 D^\ga \bD^2 \big( V D_\ga V' - V' D_\ga V \big) + \text{h.c.}
=16\big(\gO(V,V')-\gO(V',V)\big)~,
\labl{idenVV}
} 
where we have written the identity in terms of the definition
(\ref{gOgen}) of $\gO(V,V')$. Similarly, using the definition (\ref{WV5defns})
of $V_5$ it is not to difficult to show that 
\equ{
- \der_5  \gO(V, V) = \gO(V, V_5) + \gO(V_5, V) 
+ \frac 1{4 \sqrt 2} \Big( 
D^\ga( S W_\ga ) + \bD_\dga ( \bS \bW^\dga ) 
\Big)~.  
\labl{der5gO}
}

\setcounter{equation}{0}
\section{Generalization to multiple $\boldsymbol{\U{1}}$ gauge groups}
\labl{sc:MultiU1} 

The effective Lagrangians arising from Type I 
and Type II string theories generically involve several fields 
in a generalized form of the Green--Schwarz
mechanism~\cite{augusto,ibanez,antoniadis}. 
In order to describe these effective theories it is useful to generalize 
our results to multiple $\U{1}$ vector multiplets. We will asssume
multiple anomalous vector multiplets $(V^a, S^a)$, and non--anomalous
(Green--Schwarz) vector multiplets $(\cV^i, \cS^i)$. Their $\Intr_2$
parities are the same as the once given in table \ref{tb:Parities}. 
Under the $\Intr_2$ symmetry the prepotential 
of the theory must be even. The form of the prepotential
arising from generic string theories is
\be
\cP = 
\frac{1}{2} \gvf^i \gvf_i 
-  c^i_{ab}  \gvf_i \gf^a \gf^b~, 
\label{g1} 
\ee 
where we sum over all indices that appear as
upper and lower indices. The complex scalars are defined by 
\be
S^a =\frac{1}{\sqrt{2}} (\gf^a + i A_5^a)~, 
\qquad 
\cS^i =\frac{1}{\sqrt{2}} (\gvf^i + i \cA_5^i)~.
\label{g2}   
\ee
In order to work out the details of the interactions, the
Chern--Simons multiplets introduced in equations \eqref{gOgen} and 
(\ref{solY}) are generalized to  
\equ{
\widetilde \Omega^{i} = 
c^i_{ab}\Big( \Omega(V^a,V^b) -  \frac 12 V_5^a V_5^b \Big)~, ~~
Y_\ga^{i} = \frac i4 c^i_{ab} \, \bD^2  \big(
V_5^a D_\ga V^b 
\! - D_\ga V_5^a  V^b 
+ \sqrt 2 \, S^a D_\ga V^b 
\big)~. 
\labl{g3}
}
For notational convenience we have absorbed the $V_5^a V_5^b$
expression in the definition of ${\widetilde\Omega}^i$. 
Then the initial starting point of the generalized construction is the
Lagrangian
\equ{
S_{SGS} = S_{STV} + S_{Bianchi} + S_{bdy} + S_{\go^2}~, 
\labl{g4}
}
where the various parts of the action are given by
\equa{
S_{STV} & = \int  \d^5 x \bigg[ 
\int  \d^2 \gth \Big( 
\frac 14 \widehat \cW^\ga_i \widehat \cW_\ga^i 
- \frac i2 \cT_{i\, 5}^\ga 
\widehat \cW_\ga^i 
+ \frac i2 \cT^\ga_i \cW_\ga^i(\widehat \cV_5) 
\Big) + \text{h.c.} 
+  \int \d^4 \gth 
\widehat \cV_{i\, 5}^{\;} \widehat \cV^{i}_{5} 
\bigg]~, 
\labl{g5}
\\[1ex]
S_{Bianchi}  & = 
\int \d^5 x \bigg[ \int \d^2 \gth\, 
\frac 12 \cT_{i\, 5}^\ga Y_\ga^{i} 
+ \text{h.c.}  + 
\int \d^4 \gth \, 4 \cL_i {\widetilde\Omega}^{i}  
\bigg]~,   
\labl{g6} 
\\[1ex]
S_{bdy} & = \left. 
\int \d^4 x 
\int \d^2 \gth \frac i4  \gx_{a\, I}^i  \cT^\ga_i W_\ga^a (V) 
\right|_{I} + \text{h.c.}~, 
\labl{g7} 
\\[1ex]
S_{\go^2} & = \int \d^5 x \bigg[ 
\int \d^2 \gth\, \frac {1}{4} 
Y^{\ga}_i Y^{i}_{\ga} + \text{h.c.} 
- \int \d^4 \gth\,  4 {\widetilde\Omega}_i {\widetilde\Omega}^i 
\bigg]~. 
\labl{g8}
}

Following the procedure used earlier to integrate out non--dynamical fields, 
we find that in the vector multiplet formulation the generalized Lagrangian 
becomes
\equa{
S_{SGS1} &= \int \d^5 x \bigg[
\int \d^2 \gth  \Big( 
\frac 14 
\cW_i^\ga \cW^i_\ga 
- \frac i2 Y_i^\ga \cW^i_\ga \Big) 
+ \text{h.c.} + 
\int \d^4 \gth 
\Big( 
\cV_{i\, 5}
+ 4 {\widetilde\Omega}_i
\big) \cV^i_5 
\bigg] 
\non \\[1ex]
&
\quad 
+   \frac 12 \int \d^4 x  
\int \d^4 \gth 
\Big[
-2 \gx^i_{a\,I}  V^a \big( \cV_{i\,5} + 2 {\widetilde\Omega}_i  \big) 
+ \gx^i_{a\, I} \gx^{b}_{i\,I} \,  \gd(0)\,  V^a V_b  
\Big] 
\bigg|_I~. 
\labl{g9}
}
Notice in (\ref{g9}) the appearance of generalized Chern--Simons terms
whose form and transformations under the gauge variations 
\(
\gd V^a =  \Lambda^a + {\bar \Lambda}^a
\) 
and 
\(
\gd S^a = \sqrt 2 \der_5 \gL^a
\) 
are given by
\begin{eqnarray}
\label{g10a}
S_{CS} &=& -4  \gx_{c\,I}^i  c_{i\, ab}  \int \d^4 \gth \ 
\gO(V^{a},V^{b}) V^c 
\ \delta (y-I)~, \\
\delta S_{CS} &=&  \gx_{c\, I}^i c_{i\, ab} 
\int \d^2 \gth 
( \Lambda^c  W^{a\, \ga} W_\ga^b - \Lambda^a  W^{b\, \ga} W_\ga^c ) \ \delta
(y-I) \ + \ \text{h.c.}~. 
\label{g10b}
\end{eqnarray} 
As we saw in Section \ref{sc:DualGS}, this term is absent in the single 
$\U{1}_A$ case because of the identity \eqref{totSder}.  
The anomalous transformations of the classical Lagrangian (\ref{g9})
in the vector formulation are given by
\equ{ 
\arry{c}{\dsp  
\gd \cV^i = 0~, 
\quad 
\delta \cS^i = \sqrt{2} \gx_{a\, I}^i \Lambda^a \delta (y-I)~,
\\[1ex] \dsp 
\delta S_{SGS1} = -  \gx_{c\, I}^i c_{i\, ab}
\int \d^2 \gth \ 
\Lambda^a W^{b\, \ga}
W^c_\ga  \delta (y-I)  + \text{h.c.}~. 
}
\label{g11}  
}
Interestingly, the final anomalous variation (\ref{g11}) of the
action  $S_{SGS1}$ is obtained by combining the gauge variation (\ref{g10b})
of the generalized Chern--Simons term $S_{CS}$ with that of the axionic
coupling $c^i_{ab} \cS_i W^{a\, \ga}W^b_\ga$ hidden in (\ref{g9}).

Alternatively, the Lagrangian in the tensor multiplet formulation is given by
\equa{
S_{SGS2}  &  = 
\int \d^5 x \bigg[
\int \d^2 \gth \Big(  
\frac 14 \big( \cT_{5i}^\ga + Y^{\ga}_i  \big)  
\big( \cT_{\!5\,\ga}^i + Y^i_{\ga} \big) 
+  \frac i2  \gx_{a\, I}^i \cT^\ga_i W_\ga^a \,  \gd(y - I) +{\rm h.c.}\Big) 
\non \\[1ex]  & \qquad \qquad 
- \int \d^4 \gth 
\big( \cL^i - 2 {\widetilde\Omega}^i \big)\big( \cL_i - 2 {\widetilde\Omega}_i \big)
\bigg]~.
\labl{g12} 
} 
Notice again the absence of ill--defined terms in the tensor
formulation compared with the vector formulation, leading to a 
perfectly well defined description.
In the tensor formulation the corresponding gauge transformations become
\equ{ 
\arry{c}{\dsp 
\delta \cT_\ga^i = 2 i c^i_{ab} \Lambda^a  W_\ga^b~, 
\quad 
\gd \cU^i = i c^i_{ab} (\gL^a - \bgL^a) V_5^b~, 
\\[1ex] \dsp 
\delta S_{SGS2}  =
-  \gx_{c\, I}^i  c_{i\, ab}
\int \d^2 \gth \ \Lambda^a W^{b\, \ga}
W^c_\ga \delta (y-I) +  \text{h.c.}~.
}
\label{g13}
}
The dual formulations are equivalent and in particular the two 
anomalous gauge variations (\ref{g11}) and (\ref{g13}) match.
But we emphasize that in the vector formulation the generalized 
Chern--Simons term (\ref{g10a}) played a crucial role in obtaining this result.

Another interesting example where the generalized Chern--Simons terms play a
crucial role in the anomaly cancellation is the case consisting 
of only one $\Intr_2$ even $\U{1}_A$ vector multiplet and one $\Intr_2$ odd 
Green--Schwarz multiplet in the simplest case of no localized anomalies 
or Fayet--Iliopoulos terms. In this case, due to parity assignments,
the boundary gauge variation (\ref{var}) is zero. Performing a
dimensional reduction, the four dimensional Kaluza--Klein Lagrangian 
contains axionic couplings and generalized Chern--Simons terms 
(due to the five dimensional Chern--Simons terms), where both of these 
terms contain at least one massive Kaluza--Klein mode. Interestingly 
enough, the two terms are separately not gauge invariant, but their sum 
is in agreement with the original five dimensional gauge invariance of 
the model.

Finally, the supersymmetric duality relations (\ref{dualrel}) in this 
more general case take the form
\equ{
\arry{c}{  \dsp 
i \cT_{5\, \ga}^i = 
i \big( \der_5 \cT_\ga^i + \cW_\ga^i (\cU) \big) = 
\cW_\ga^i (\cV) -  i  Y_\ga^{i}~, 
\\[1ex] \dsp 
\cL^i = 
 \frac i4  \big( D^\ga\cT_\ga^i - \bD_\dga \bcT^{i\, \dga} \big) = 
\cV_5^i + 2 {\widetilde\Omega}^i - \gx_{a\, I}^i V^a \gd(y-I)~. 
}\labl{g14}
}

\end{document}